\documentclass[preprint]{aastex701}

\usepackage{natbib}
\usepackage{amsmath}

\usepackage[version=4]{mhchem}

\newcommand{\kms}{\mbox{km\,s$^{-1}$}}

\newcommand{\Msun}{\mbox{M$_{\odot}$}}

\newcommand{\E}[1]{\hbox{$10^{ #1 }$}}

\begin{document}

\title{FIRST RESULTS FROM ALPPS: A SUB-Alfv\'enic STREAMER IN SVS13A}
\shortauthors{Cort\'es et al.}

\author[0000-0002-3583-780X]{P. C. Cort\'es}
\affiliation{Joint ALMA Observatory, Alonso de C\'ordova 3107, Vitacura, Santiago, Chile}
\affiliation{National Radio Astronomy Observatory, 520 Edgemont Road, Charlottesville, VA 22903, USA}
\email{paulo.cortes@alma.cl}

\author[0000-0002-3972-1978]{J. E. Pineda}
\affiliation{Max-Planck-Institut f\"ur extraterrestrische Physik, Giessenbachstrasse 1, D-85748 Garching, Germany}
\email{jpineda@mpe.mpg.de}

\author[0000-0002-5507-5697]{T.-H. Hsieh}
\affiliation{Max-Planck-Institut f\"ur extraterrestrische Physik, Giessenbachstrasse 1, D-85748 Garching, Germany}
\affiliation{Taiwan Astronomical Research Alliance (TARA), Taiwan; Institute of Astronomy and Astrophysics, Academia Sinica, PO Box 23-141, Taipei, 106, Taiwan}
\email{thhsieh@mpe.mpg.de}

\author[0000-0002-6195-0152]{J. J. Tobin}
\affiliation{National Radio Astronomy Observatory, 520 Edgemont Road, Charlottesville, VA 22903, USA}
\email{jtobin@nrao.edu}

\author[0000-0002-0028-1354]{P. Saha}
\affiliation{Academia Sinica, PO Box 23-141, Taipei, 106, Taiwan}
\email{s.piyali16@gmail.com}

\author[0000-0002-3829-5591]{J. M. Girart}
\affiliation{Institut de Ci\'encies de l'Espai (ICE-CSIC), Campus UAB, Carrer de Can Magrans S/N, E-08193 Cerdanyola del Vall\'es, Catalonia}
\affiliation{Institut d'Estudis Espacials de Catalunya (IEEC), E-08034 Barcelona, Catalonia}
\email{girart@ieec.cat}

\author[0000-0002-5714-799X]{V. J. M. Le Gouellec}
\affiliation{Institut de Ci\'encies de l'Espai (ICE-CSIC), Campus UAB, Carrer de Can Magrans S/N, E-08193 Cerdanyola del Vall\'es, Catalonia}
\affiliation{Institut d'Estudis Espacials de Catalunya (IEEC), E-08034 Barcelona, Catalonia}
\email{valentin.le.gouellec@gmail.com}

\author[0000-0003-3017-4418]{I. W. Stephens}
\affiliation{Department of Earth, Environment and Physics, Worcester State University, Worcester, MA 01602, USA}
\email{iwstephens@gmail.com}

\author[0000-0002-4540-6587]{L. W. Looney}
\affiliation{Department of Astronomy, University of Illinois, 1002 W Green St., Urbana, IL 61801, USA}
\email{lwl@illinois.edu}

\author{E. Koumpia}
\affiliation{Joint ALMA Observatory, Alonso de C\'ordova 3107, Vitacura, Santiago, Chile}
\affiliation{European Southern Observatory, Alonso de C\'ordova 3107, Vitacura, Santiago, Chile}
\email{Evgenia.Koumpia@eso.org}

\author[0000-0002-0347-3837]{M. T. Valdivia-Mena}
\affiliation{European Southern Observatory, Karl-Schwarzschild-Strasse 2, 85748 Garching, Germany}
\email{MariaTeresa.ValdiviaMena@eso.org}

\author[0000-0001-8266-0894]{L. Cacciapuoti}
\affiliation{Joint ALMA Observatory, Alonso de C\'ordova 3107, Vitacura, Santiago, Chile}
\affiliation{European Southern Observatory, Alonso de C\'ordova 3107, Vitacura, Santiago, Chile}
\email{luca.cacciapuoti@eso.org}

\author[0000-0002-8120-1765]{C. Gieser}
\affiliation{Max-Planck-Institut f\"{u}r Astronomie, K\"{o}nigstuhl 17, D-69117 Heidelberg, Germany}
\email{gieser@mpia.de}

\author[0000-0003-1252-9916]{S. S. R. Offner}
\affiliation{Department of Astronomy, University of Texas at Austin, TX 78712, USA}
\email{soffner@utexas.edu}

\author[0000-0002-3972-1978]{P. Caselli}
\affiliation{Max-Planck-Institut f\"ur extraterrestrische Physik, Giessenbachstrasse 1, D-85748 Garching, Germany}
\email{caselli@mpe.mpg.de}

\author[0000-0002-7125-7685]{P. Sanhueza}
\affiliation{Department of Astronomy, School of Science, The University of Tokyo, 7-3-1, Hongo, Bunkyo-ku, Tokyo 113-0033, Japan}
\email{patosanhueza@gmail.com}

\author[0000-0003-3172-6763]{D. Segura-Cox}
\affiliation{Department of Physics and Astronomy, University of Rochester, Rochester, NY 14627-0171, USA}
\email{d.segura-cox@rochester.edu}

\author[0000-0001-5811-0454]{M. Fern\'andez-L\'opez}
\affiliation{Instituto Argentino de Radioastronom\'\i a (CCT-La Plata, CONICET; CICPBA; UNLP), C.C. No. 5, 1894, Villa Elisa, Buenos Aires, Argentina}
\email{manferna@gmail.com}

\author[0000-0002-6752-6061]{K. Morii}
\affiliation{Harvard-Smithsonian Center for Astrophysics, 60 Garden Street, Cambridge, MA 02138, USA}
\email{kaho.morii@cfa.harvard.edu}

\author[0000-0001-7393-8583]{B. Huang}
\affiliation{Institut de Ci\'encies de l'Espai (ICE-CSIC), Campus UAB, Carrer de Can Magrans S/N, E-08193 Cerdanyola del Vall\'es, Catalonia}
\affiliation{Institut d'Estudis Espacials de Catalunya (IEEC), E-08034 Barcelona, Catalonia}
\email{hb6170@gmail.com}

\author[0000-0002-7945-064X]{F. O. Alves}
\affiliation{Institut de Radioastronomie Millim\'etrique (IRAM), 300 rue de la Piscine, F-38406, Saint-Martin d'H\`eres, France}
\email{alves@iram.fr}

\author[0000-0003-2384-6589]{Q. Zhang}
\affiliation{Harvard-Smithsonian Center for Astrophysics, 60 Garden Street, Cambridge, MA 02138, USA}
\email{qzhang@cfa.harvard.edu}

\author[0000-0003-4022-4132]{W. Kwon}
\affiliation{Department of Earth Science Education, Seoul National University, 1 Gwanak-ro, Gwanak-gu, Seoul 08826, Republic of Korea}
\affiliation{SNU Astronomy Research Center, Seoul National University, 1 Gwanak-ro, Gwanak-gu, Seoul 08826, Republic of Korea}
\email{wkwon@snu.ac.kr}

\author[0000-0002-8975-7573]{C. L. H. Hull}
\affiliation{AES Indiana, 1 Monument Circle, Indianapolis, IN 46204, USA}
\email{chat.hull.work@gmail.com}

\author[0000-0002-7402-6487]{Z.--Y. Li}
\affiliation{Department of Astronomy, University of Virginia, 530 McCormick Rd., Charlottesville, VA 22904, USA}
\email{zl4h@virginia.edu}

\begin{abstract}
We present the first results from the ALMA Perseus Polarization Survey (ALPPS), focusing on the magnetic field in the SVS13A circumbinary disk. The dataset includes full-Stokes dust continuum observations at $\sim0\farcs3$ and 870 $\mu$m, as well as molecular line emission from C$^{17}$O$(J=3 \rightarrow 2)$ at $\sim0\farcs3$, C$^{18}$O$(J=2 \rightarrow 1)$ at $\sim0\farcs2$, and DCN$(J=3 \rightarrow 2)$ at $\sim0\farcs1$ angular resolution. Our observations resolve both a previously identified dust spiral and an infalling streamer, capturing their spatial and kinematic structures. The streamer is traced from scales $>300$ au down to the circumbinary disk.
Using alignment measure (AM) maps and histograms that compare the orientations of the plane-of-sky magnetic field with local intensity and velocity gradients, we find that the AM distribution peaks at a value of 1. This AM peak strongly suggests alignment between the field and the dust total intensity emission, as well as between the field and the gas velocity, which in turn suggests grain alignment by magnetic fields.
From our data, we derive a magnetic field strength, B$_{\mathrm{pos}} \sim 1.1 \pm 0.6$\, mG, and a kinetic to magnetic energy ratio of $0.5 \pm 0.4$, suggesting magnetic dominance. We also produced a map of the Alfvénic Mach number, finding $\mathcal{M}_{\rm A} < 1$ along the streamer, consistent with sub-Alfvénic infalling motions.
Therefore, the field is likely facilitating the inflow of material from the envelope onto the disk by constraining movement across the field lines. This represents the first detection of a magnetically sub-Alfv\'enic infalling streamer in a protostellar system.
\end{abstract}


\section{INTRODUCTION}\label{se:intro}
\setcounter{footnote}{1}
Stars form within dense, filamentary, and weakly ionized gas and dust complexes commonly known as molecular clouds \citep{Pineda2023}.
Despite their weak ionization rates, magnetic fields thread these regions, and their effects are unavoidable. 
In recent decades, the advent of millimeter and submillimeter interferometers such as the Berkeley, Illinois, and Maryland Association (BIMA), the Combined Array for Research in Millimeter-wave Astronomy (CARMA), and Sub-Millimeter Array (SMA) opened the window for the study of magnetic fields in star-forming regions through the detection of linearly polarized dust emission at increasingly high sensitivities and ever smaller angular scales \citep{Girart1999a, Cortes2005, Girart2006, Hull2014, Zhang2014, Galametz2018}.
The arrival of the Atacama Large millimeter/sub-millimeter Array (ALMA), the Stratospheric Observatory for Infrared Astronomy (SOFIA) \citep{Harper2018}, and the James Clerk-Maxwell Telescope JCMT-POL2 \citep{Ward-Thompson2017} further enhances the potential for large surveys to investigate the role of magnetic fields in the formation of stars \citep{Pattle2023}. To date, a complete understanding of their effects remains elusive. One area where this is more evident is the
scales associated with the infalling and streaming of material onto disks around protostars ($ < 100 $ au), where observations of magnetic fields are not abundant. For instance, \citet{Cox2018}  mapped 10 Young Stellar Objects (YSO) in Perseus in Full Stokes with ALMA at a resolution of 80 au. Also, \citet{Sadavoy2018b,Sadavoy2019} mapped approximately 37 YSOs in Ophiuchus in full Stokes with ALMA at a resolution of 30 au, detecting polarized emission in only 14 cases. Although most of Sadavoy et al.'s  detections (9) are consistent with self-scattering, the remaining detections are associated with magnetic fields whose morphologies are either poloidal or ``hourglass'' shaped, ruling out toroidal configurations.  Unfortunately, both studies included only dust continuum, so there was no kinematical information to correlate possible streamers with the field at the same scales. 

 Since their discovery, streamers have significantly altered the classic idea of monolithic collapse in star formation \citep{Pineda2023}.
Streamers have been seen in a number of protostellar systems \citep{Pineda2020,Ginski2021,Murillo2022,Valdivia-Mena2022,Valdivia-Mena2023,Hsieh2023,Hsieh2024}, suggesting that the mechanism under which stars acquire their mass is more dynamic than the classic paradigm. Here, we separate infalling from accreting, where the former corresponds to material falling into the potential well of the protostar from the envelope and the latter from the disk onto the star.
These infalling streamers have the potential to continuously feed the disk and the envelope, which can change the evolutionary stage of the system, including its time-scales and final mass \citep{Kuffmeier2024}. 
Furthermore,  streamers resemble for low-mass star formation the competitive accretion mechanism proposed for high-mass star formation in environments with significant fragmentation. In the latter, more massive protostars dominate their environments, allowing them to accrete more efficiently from the shared gas reservoir than their lower-mass neighbors \citep{Bonnell2001,Beuther2025}.

To advance characterization of the physics of star formation, we introduce here the ALPPS project,  a survey of 38 YSOs in the Perseus molecular complex in all Stokes parameters with ALMA. This survey samples the magnetic field through the detection of polarized dust emission at scales of $0^{\prime\prime}.3$ along with the gas kinematics by observing a number of molecular lines. From this survey,  the SVS13A complex, a close binary system composed of two sources, VLA 4A and VLA 4B \citep{Anglada2000}, surrounded by a circumbinary disk \citep{Tobin2018,Bianchi2022,Diaz-Rodriguez2022}, stands out as a remarkable case. SVS13A is located in the Perseus region at a distance of $d= 300$ pc \citep{Ortiz-Leon2018}, making the ALPPS resolution about 60 au. This strong millimeter source has been suggested as the origin of the  HH 7-11 outflow system, where an [Fe {\small II}], $0^{\prime\prime}.2$ long jet was discovered extending from VLA 4A in the direction of HH 7-11 system \citep{Hodapp2014}.  
Recently, \citet{Diaz-Rodriguez2022} observed SVS13A with ALMA at 30 au resolution in Band 7. Their data resolved the binary system, confirming a remarkable spiral feature in the dust emission previously seen by \citet{Tobin2018}. They resolved the structure into two spiral arms extending up to approximately 375 au and an estimated total mass of around 0.052 {\Msun}. By using rotational diagrams from \ce{CH3OCHO} and \ce{CH2DOH}, they obtained temperatures of about 140 K in a small region of 60 au around VLA 4A. Similarly, values of 90 K and 160 K were determined around VLA 4B, although those are considered to be less reliable due to increased dust opacity \citep{Diaz-Rodriguez2022}. For the circumbinary disk, the dust emission appeared to be optically thin, and no reliable temperature estimate was obtained. Nonetheless, they assumed a constant value of 140 K, which was used to derive the circumbinary disk mass.

The paper is outlined as follows, Section \ref{se:obs} introduces the observing setup, Section \ref{se:res} the results, Section \ref{se:discussion} the discussion, Section \ref{se:conclusion} presents the summary and conclusions, while Appendices \ref{se:appendix_a}, \ref{sse:am}, and \ref{sse:small_window} contain supplemental material supporting the interpretation of the data.

\section{OBSERVATIONS}\label{se:obs}

SVS13A was observed as part of project 2021.1.00418.S, which was executed in session mode  \citep[see chapter 8 in ][ for details about the session observing mode]{Cortes2023} from August 21$^{\mathrm{st}}$ 2022 to September 14$^{\mathrm{th}}$ 2022 in configuration C43-4 providing baseline lengths from 15 m to 1301 m, project 2017.1.00053.S executed from December 17$^{\mathrm{th}}$ 2017 and September 29$^{\mathrm{th}}$ 2018 in configuration C43-5 providing baselines between 43 m and 1400 m, and project 2022.1.00479.S observed 
between June 10th 2023 and June 23rd 2023 in configuration C43-7 providing baselines between 64 m and 6300 m. 
For the former project,
the correlator was configured to yield full polarization cross correlations ($XX, XY, YX,$  and $YY$) using the Frequency Division Mode  (FDM), including spectral windows to map the dust continuum and windows centered on a number of molecular line rotational transitions relevant to the study of star formation, like $^{12}$CO$(J=3 \rightarrow 2)$ configured at higher spectral resolution to resolve the outflows  (see Table \ref{table:setup}). The observations were obtained at an effective wavelength of 0.88 mm. 
The bandpass was calibrated using  J0423-0120, J0510+1800, and J0238+1636. The time-dependent gain and the instrumental polarization terms were calibrated using J0336+3218 and  J0522-3627, respectively.
For calibration and imaging, we used CASA versions 5.4 and 6.5, respectively \citep{CASA2022}.
To image the continuum, we manually extracted the line-free channels from each spectral window,
and we later iteratively self-calibrated the phase of these data using a final solution interval of 30 seconds ($\sigma_{\mathrm{rms}} = 200\,\, \mu\mathrm{Jy\,beam}^{-1}$ with robust=0.5). 
These solutions were then applied to all of the molecular line transitions presented here before imaging. The spectral cubes were produced by using a 1 {\kms} channel width.  
The Stokes parameters were imaged using the CASA task {\em tclean}, which yielded an approximate angular resolution of  $0.4^{\prime \prime} \times 0.3^{\prime \prime}$, with an approximate position angle of $27^{\circ}$. We covered the baseline weighting parameter space from -2 to 2 in steps of 0.5 to explore what structures are recovered in Stokes space for both line and continuum, except for DCN (see below). The values used for each map are stated in the corresponding figure caption.
The DCN$(J=3\rightarrow2)$ data were obtained from project 2022.1.00479.S where the visibilities were calibrated and imaged using CASA 6.4.12.
The imaging  is made with a robust weighting of 0.5 and 2, resulting in beam sizes of $0\farcs11\times0\farcs08$ ($\sim$30 au) and
$0\farcs2\times0\farcs15$ ($\sim$ 45 au), respectively.

\setlength{\tabcolsep}{2pt}
\begin{deluxetable*}{cccccccccccccccccc}
\tablecolumns{12}
\tablewidth{0pt}
\tabletypesize{\scriptsize}
\tablecaption{ Spectral Setup\label{table:setup}}
\tablehead{
\colhead{Band} &
\colhead{Spw 1} &
\colhead{$\Delta \nu_{1}$} &
\colhead{Spw 2} &
\colhead{$\Delta \nu_{2}$} &
\colhead{Spw 3} &
\colhead{$\Delta \nu_{3}$} &
\colhead{Spw 4} &
\colhead{$\Delta \nu_{7}$} &
\colhead{$B_\textrm{maj}$} &
\colhead{$B_\textrm{min}$} &
\colhead{PA} \\
\colhead{}     &
\colhead{(GHz)}     &
\colhead{(kHz)}     &
\colhead{(GHz)}     &
\colhead{(kHz)}     &
\colhead{(GHz)}     &
\colhead{(kHz)}     &
\colhead{(GHz)}     &
\colhead{(kHz)}     &
\colhead{($^{\prime \prime}$)}  &
\colhead{($^{\prime \prime}$)}  &
\colhead{(\arcdeg)}
}
\startdata
7 & 334.5 & 976.562 & 336.5 & 976.562 & 345.8  & 244.141 & 348.5 & 976.562 &  0.35 & 0.27 & -26.5 \\
\enddata
\vspace{0.1cm}
\tablecomments{The spectral configuration and beam size from the ALPPS project is detailed here. For each spectral window (e.g., Spw 1), the center frequency is indicated in GHz, along with the channel width in kHz (e.g., $\Delta \nu_{1}$).  The  synthesized beam parameters (angular resolution) and the major axis $B_\textrm{maj}$, minor axis $B_\textrm{min}$, and position angle PA are listed from images using robust=0.5}
\end{deluxetable*}

\section{RESULTS}
\label{se:res}

\begin{figure*} 
\includegraphics[width=0.95\hsize]{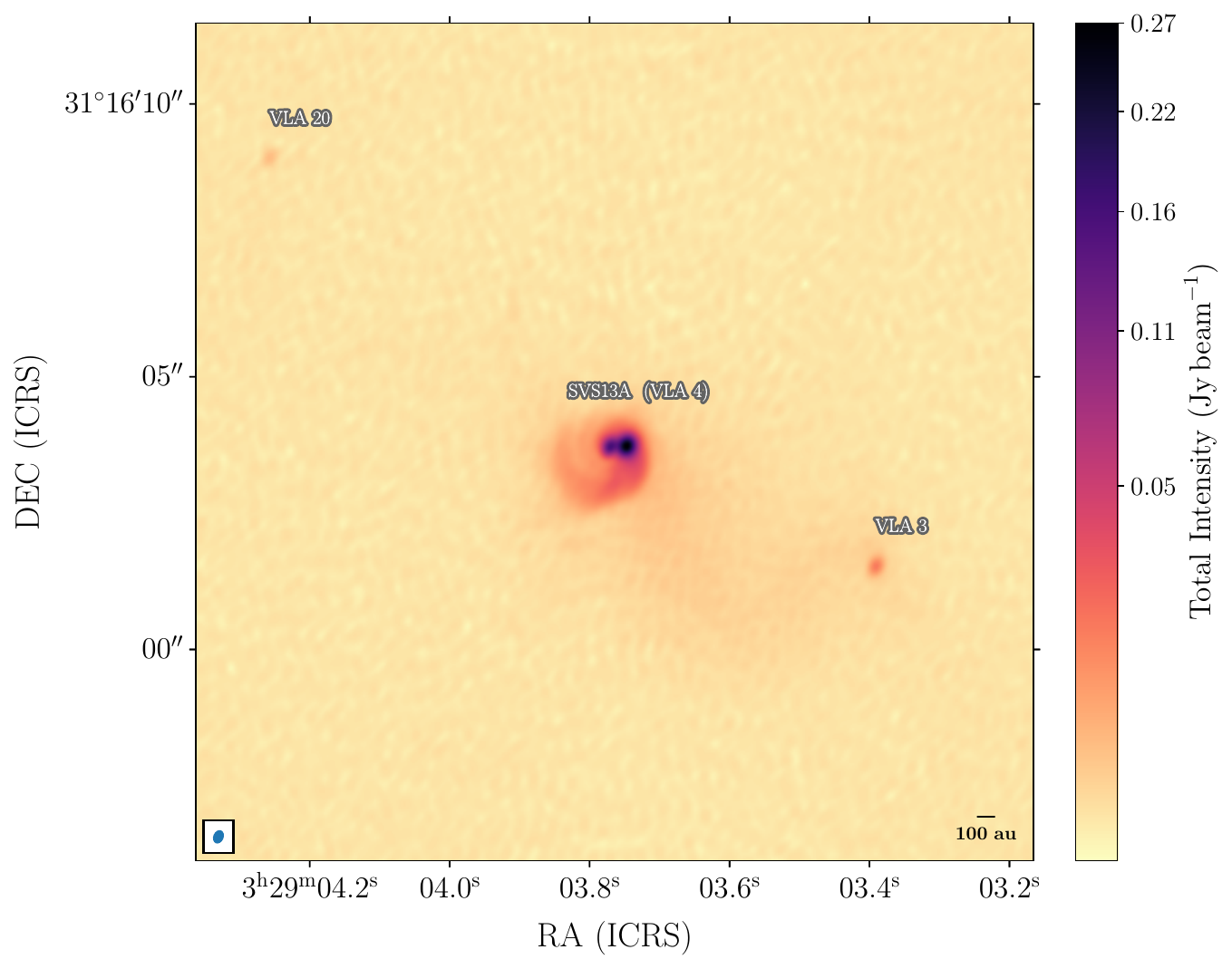}
\smallskip
\caption{Dust continuum emission maps from the SVS13A data shown in color scale. The data 
 were imaged using a robust weight parameter of -0.5 to highlight the resolved VLA 4 binary system and nearby sources, which are indicated by the labels. The relative size of the beam is indicated as a blue ellipse in the bottom left corner of the map, and a scale bar corresponding to 100 au length is indicated in the bottom right corner of the map.
\label{fig:dust_map}
}
\bigskip
\end{figure*}

\subsection{Dust Continuum Emission}

\begin{figure*} 
\centering

\includegraphics[width=0.75\hsize]{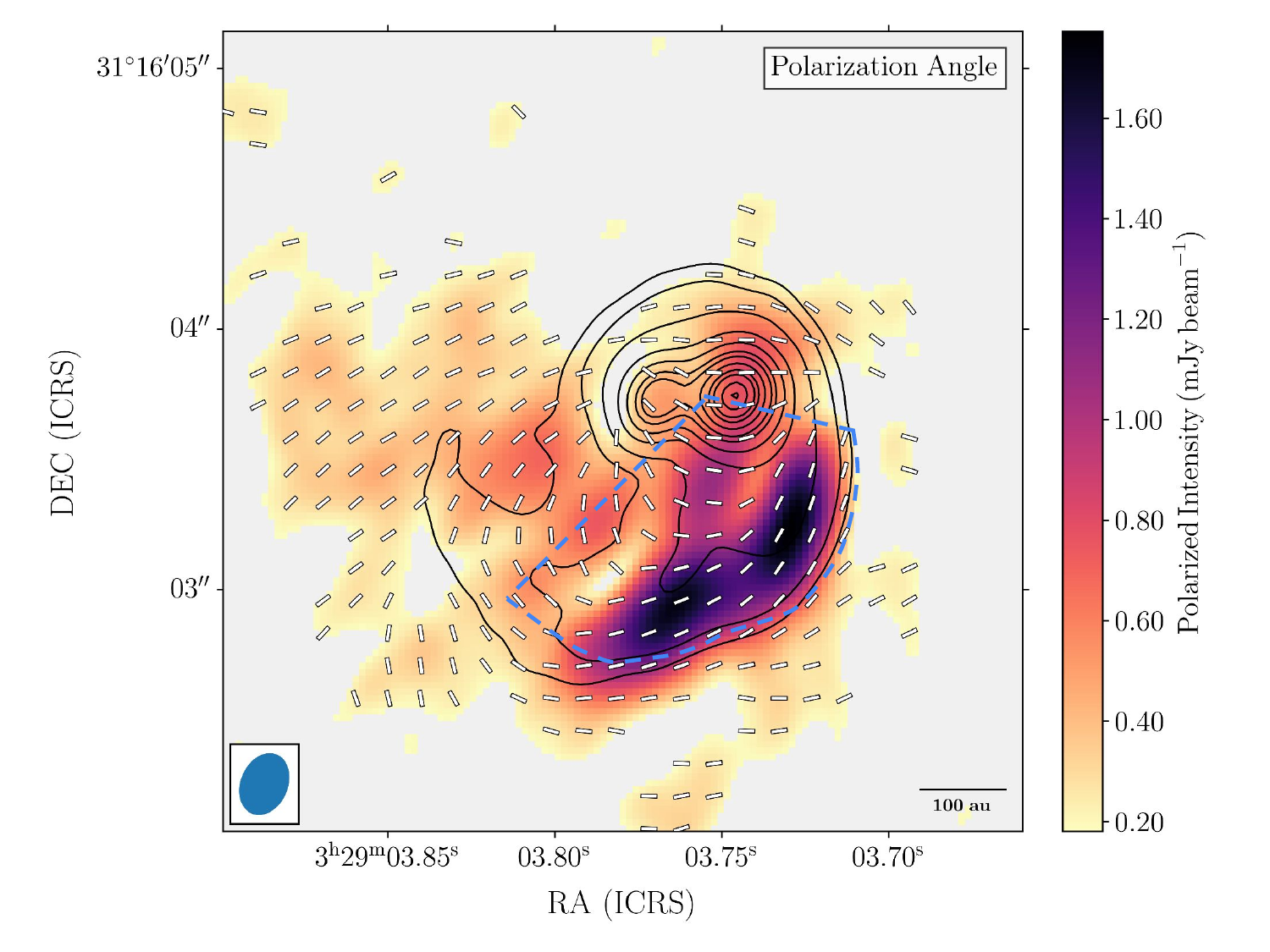}\\
\includegraphics[width=0.75\hsize]{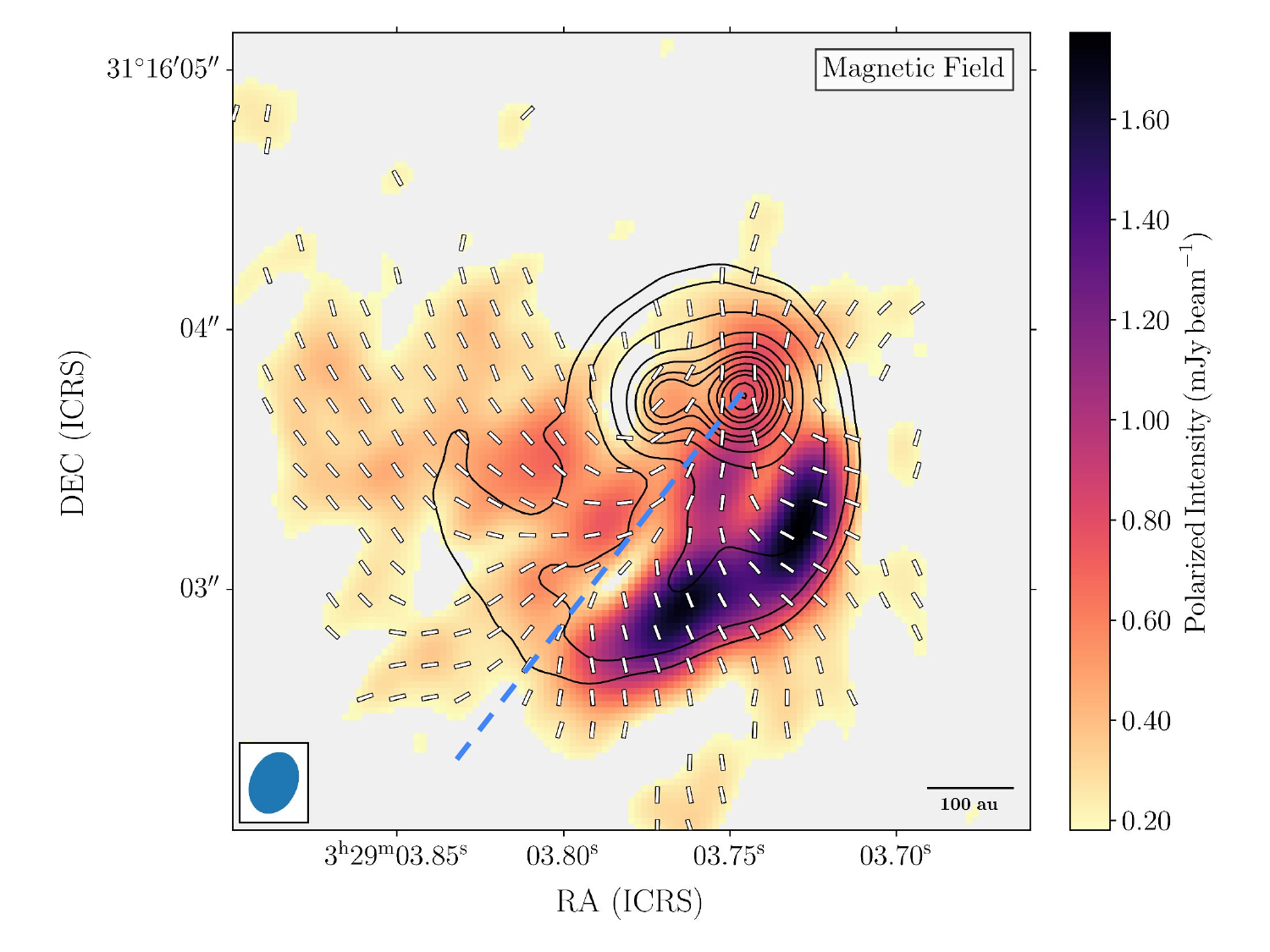}\\
\smallskip
\caption{{\bf\em Upper panel.} The polarization position angle map from the SVS13A polarized dust continuum emission as white pseudo vectors. In color scale, we show the polarized intensity emission imaged using robust=0.5, and in contours, we show the total intensity dust continuum imaged with robust=-2 and using levels of 10,  15,  28.7,  54.9,  81.1, 110, 130, 160, 190, 210, 240, and 260 mJy beam$^{-1}$. The blue oval indicates the beam obtained by imaging the data with robust=0.5. The blue segmented lines indicate the region where the polarization position angle appears to follow the spiral substructure seen in the dust emission. {\bf\em Lower panel.} Following the upper panel, the inferred magnetic field morphology map is shown under the assumption of grain alignment by magnetic fields (i.e., pseudo vectors rotated by 90$^{\circ}$). The blue segmented line indicates the region where the inferred magnetic field lines appear to bifurcate.
\label{fig:pol_map}
}
\bigskip
\end{figure*}

Figure~\ref{fig:dust_map} presents the 880\,$\mu$m continuum emission map of SVS13A obtained with ALMA. The sources VLA~3, SVS13A (VLA~4), and VLA~20 are  detected and in agreement with the 1\,mm and 0.9\,mm continuum maps reported by \citet{Tobin2018} and \citet{Diaz-Rodriguez2022}, respectively. The two main components of the SVS13A binary system, VLA~4A and VLA~4B, are spatially resolved, along with the prominent spiral structure in the surrounding dust.
To the southwest of SVS13A, a faint bridge of continuum emission is observed connecting the binary system with the VLA~3 protostellar core\footnote{The data presented in Figure \ref{fig:pol_map} were imaged with robust=-2 to resolve the binary, which makes the bridge tenuous. When imaging with robust=0.5, more emission is recovered and the bridge appears with more significance.}. This feature is not evident in the \citet{Diaz-Rodriguez2022} map (see their Figure~6). Our data also encompass the more isolated VLA~20 core, which is detected at a lower signal-to-noise ratio (S/N).
The prominent spiral substructure seen in our continuum map is consistent with the higher resolution results of \citet{Tobin2018} and  \citet{Diaz-Rodriguez2022}, showing two spiral arms with the same orientation. While spiral patterns are commonly observed in protoplanetary disks—such as those around Elias~27, IM~Lup, and WaOph~6 \citep{Huang2018}—these are typically symmetric around a single central star. In contrast, the spiral substructure in SVS13A has a single arm emerging from a circumbinary disk around a close binary system.
Although single spiral substructures are not common, we note that spiral structures are frequently reproduced in magnetohydrodynamic (MHD) simulations, where they are often attributed to gravitational instabilities triggered by dynamical perturbations, such as stellar flyby events \citep{Tokuda2014,Matsumoto2015,Lebreuilly2024}.

\subsection{The polarized dust emission}
\label{sse:pmorph}

Figure \ref{fig:pol_map} shows both the polarization position angle map and the inferred magnetic field morphology map, projected onto the plane of the sky under the assumption of magnetic alignment of dust grains (i.e., the polarization position angle is rotated by 90$^{\circ}$). To produce the polarization maps, we debiased the Stokes $Q$ and $U$ images pixel by pixel, enforcing a 3\,$\sigma$ cutoff from Stokes $I$ (total intensity), by following \citet{Vaillancourt2006} and \citet{Hull2015}.
From visual inspection, the polarized position angle map (upper panel in Figure \ref{fig:pol_map}) shows axisymmetric features around the binary system that appear consistent with the spiral structure seen in the dust emission up to a radius of $\sim$ 0$^{\prime\prime}$.8, or $\sim 240$ au  (the region is indicated by blue segmented lines), with the distance taken from the mid-point between the VLA 4A and 4B cores. Beyond this point to the east, the polarization morphology becomes mostly uniform in a north-west to south-east direction. 
On the other hand, the inferred magnetic field morphology (lower panel in Figure \ref{fig:pol_map}), shows orientations which we divide into three main components. Over the binary, the field appears pinched, resembling an ``hourglass'' shape aligned with the minor axis of the binary system. To the east of the binary, the pinched morphology evolves into a clear spiral pattern that extends beyond the 240 au radius. To the south, the field bifurcates into a more uniform pattern, which also extends beyond the lowest dust emission contour.  We indicate the location of the field bifurcation by a dashed blue line in Figure \ref{fig:pol_map}, lower panel.
Note that the lowest dust emission contour shown in the map has an S/N of 40\,$\sigma$ with $\sigma=0.25$ mJy\,beam$^{-1}$. 
It is noteworthy that the polarized emission varies smoothly and consistently across the region, without abrupt changes or discontinuities, indicating that a coherent structure exists on the scale of the circumbinary disk ($\sim 45$ au). Furthermore, 
the polarized emission also extends well into the dust bridge that connects the circumbinary disk with the VLA 3 protostar. We will be exploring that connection in an upcoming work.

\subsubsection{Polarization Fraction}
\label{sse:pfrac}

\begin{figure*} 
\centering
\includegraphics[width=0.99\hsize]{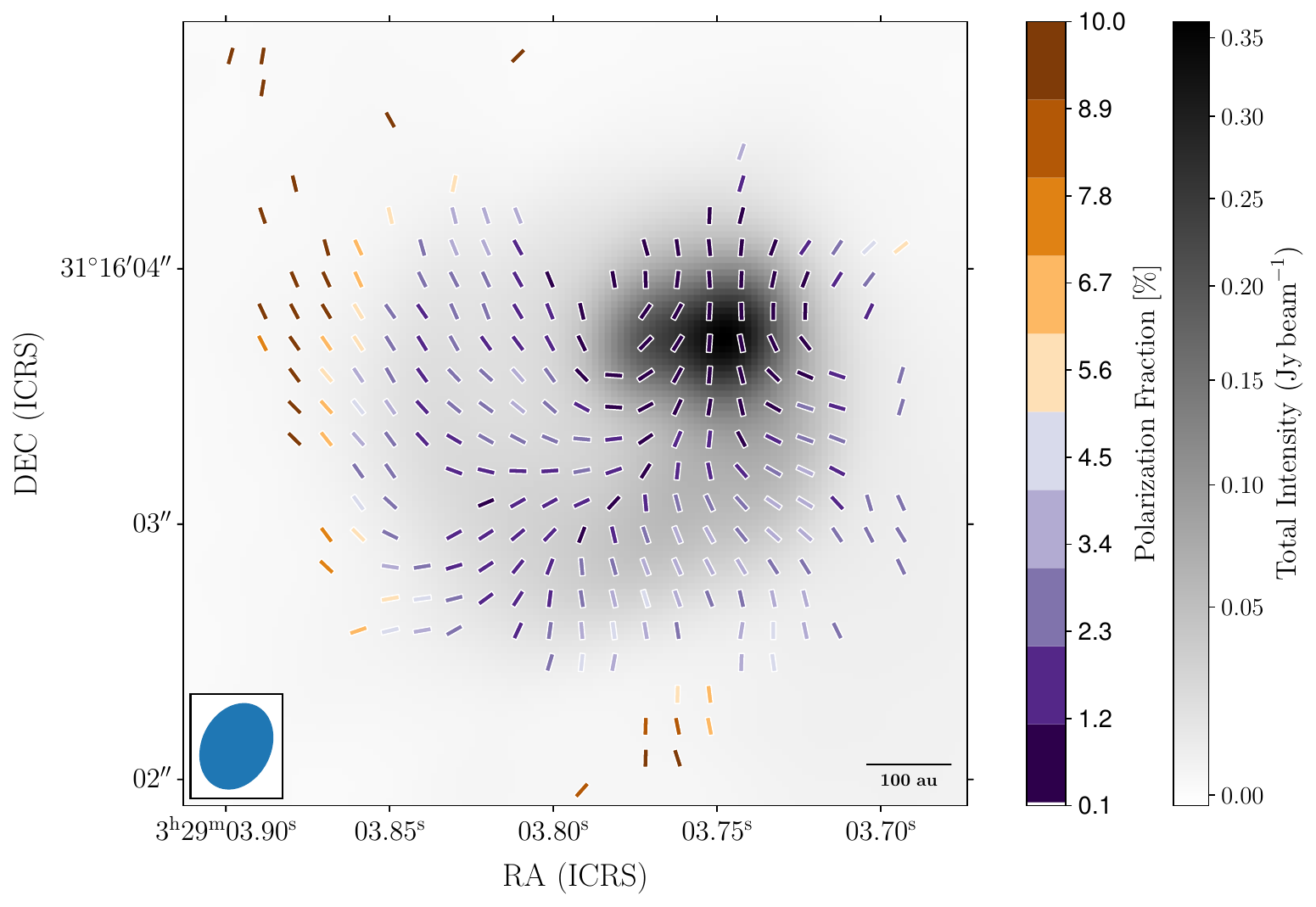}
\smallskip
\caption{The polarization map of SVS13A is shown in gray-scale using robust=0.5. The fractional polarization is shown as a pseudo-vector map where each pseudo-vector color indicates a fractional polarization level, indicated by the colorbar. The beam is shown as a blue ellipse in the bottom left corner, and the scale is highlighted in the bottom right corner. 
\label{fig:pfrac}
}
\bigskip
\end{figure*}

Figure \ref{fig:pfrac} shows the polarization fraction over the SVS13A circumbinary disk. Although the fractional polarization ranges from 0.15\% to 30\%, we put a cap of 10\% in the colorbar to allow for a better visualization of the low fractional polarization values throughout the disk.  The higher values $(> 10\%)$ are seen outside the circumbinary disk and might be the result of missing flux in the total intensity emission due to the angular scales sampled by ALMA in the C-43 configuration. Over the binary system, the fractional polarization ranges between 0.15\%, marginally close to the ALMA accuracy \citep{Cortes2023}, and $\sim 1\%$. As we move outside the binary system, the fractional polarization increases, but stays below 10\% throughout most of the dust spiral substructure. The observed range of fractional polarization values is consistent with results from other regions in Perseus as presented by \citet{Cox2018} and in Ophiuchus at higher angular resolution \citep{Sadavoy2018a}. 

\bigskip
\bigskip
\subsection{Molecular emission}
\label{sse:molecules}

\subsubsection{$C^{17}O\,(J=3\rightarrow2)$ }
\begin{figure*} 
\centering
\includegraphics[width=0.99\hsize]{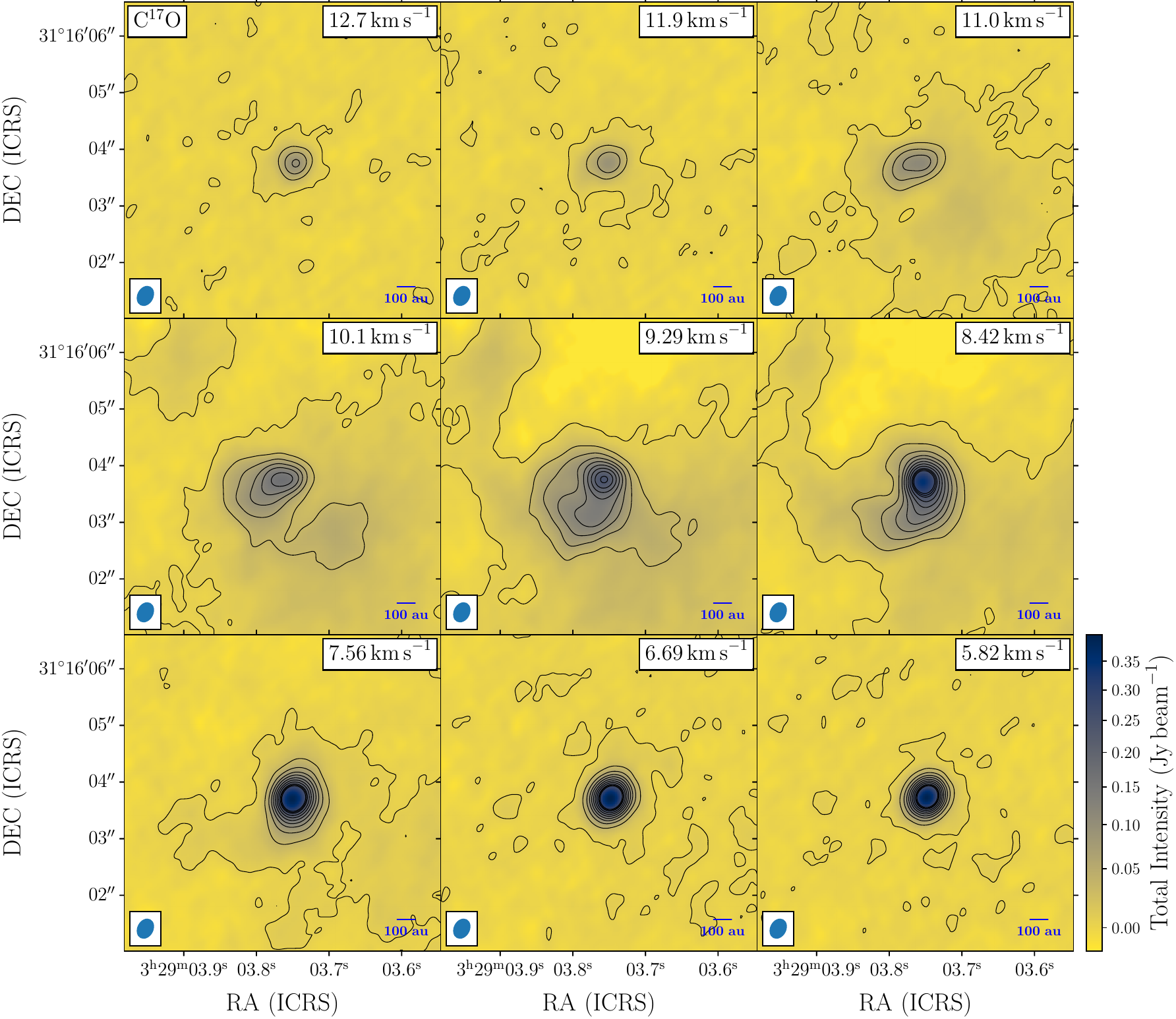}
\caption{Channel maps of C$^{17}$O emission in color scale between 5.8 {\kms} and 12.7 \kms\ with the channel velocity indicated in white at the top right corner within each channel map. Superposed are C$^{17}$O contours in black with values 4.32,  19.1,  33.8,  48.6,  63.4,  78.1,  92.9, 110,120, 140, 150, 170, 180, 200, 210, 230, 240, 260, and 270 mJy beam$^{-1}$. The beam is represented as a blue ellipse in the bottom left corner,  and the scale is shown in the bottom right corner. 
\label{fig:c17o_chan_map}
}
\end{figure*}

The spectral setup of the ALPPS project was configured to observe both the $^{12}$CO$(J=3 \rightarrow 2)$ and C$^{17}$O$(J=3 \rightarrow 2)$ rotational transition lines simultaneously. While the former usually traces outflow emission, which we are not exploring here, the latter appears to trace the gas motions throughout the circumbinary disk. In fact, it appears to also trace the spiral substructure seen in the dust continuum emission. 
Figure \ref{fig:c17o_chan_map} shows the C$^{17}$O velocity channel maps while Figure \ref{fig:c17o} shows the integrated velocity (moment 0) and velocity field (moment 1) maps of C$^{17}$O emission superposed with dust emission contours. From the channel maps, the spiral substructure becomes apparent when inspecting the 7.56 {\kms} and 8.42 \kms\ channel maps. Additional weaker emission is seen to extend to the south-west up to the VLA 3 proto-stellar core, consistent with the dust bridge previously mentioned.  The C$^{17}$O moment 0 map confirms that the extent of the gas emission is similar to that of continuum emission when using a robust equal to 0.5 and the same statistical threshold, or $5\,\sigma_{I}$ with $\sigma_{I} = 2\times10^{-4}$ Jy\,beam$^{-1}$. When inspecting the moment 1 map, the spiral structure is also seen as a gradient from $-0.6$ \kms\, to 1.0 {\kms}, where the systemic velocity of the VLA 4A protostellar core is $v_{\mathrm{lsr}} = 7.36$\,\, {\kms}\, \citep{Diaz-Rodriguez2022}. 
Furthermore, when inspecting the C$^{17}$O velocity channels, gas structures connect with the spiral from 10.1 {\kms} to 8.42 {\kms}, suggesting that C$^{17}$O might also be tracing streaming gas infalling onto the circumbinary disk as previously shown by \citet{Hsieh2024}.
The  C$^{17}$O emission appears single-peaked through the whole region with no evidence of self-absorption, which suggests optically thin emission. Thus, the moment 1 map can be considered to be a good representation of the gas motions through the line of sight. 

The gas kinematics on SVS13A was studied by \citet{Diaz-Rodriguez2022} who concluded that most of gas emission appears to show rotational signatures around the binary system. \citet{Hsieh2024} also mapped SVS13A using NOEMA, where their CH$_{3}$CN results are  consistent with a rotating envelope around
VLA 4A rather than VLA 4B or the circumbinary disk. This behavior is expected from complex organic molecules such as CH$_{3}$CN whose emission will trace the regions close to the binary (hot corinos) and not the extended circumbinary disk because of the high sublimation temperature (T $\sim$ 100 K) of complex organic molecules \citep{Collings2004}. Also, \citet{Diaz-Rodriguez2022}'s high-resolution data did not sample structures extending beyond the binary. Although our C$^{17}$O data can trace the spiral in the gas, where the morphology of the moment 1 map does not suggest rotation at the scales of the circumbinary disk,
the velocity resolution in our data ($\sim 0.9$ \kms)  is not sufficient to trace the velocity gradient effectively.  We do this instead by using C$^{18}$O$(J=2\rightarrow1)$ emission (see section \ref{ssse:c18o}).

\subsubsection{$C^{18}O\,(J=2\rightarrow1)$}
\label{ssse:c18o}

\begin{figure*} 
\centering
\includegraphics[width=0.49\hsize]{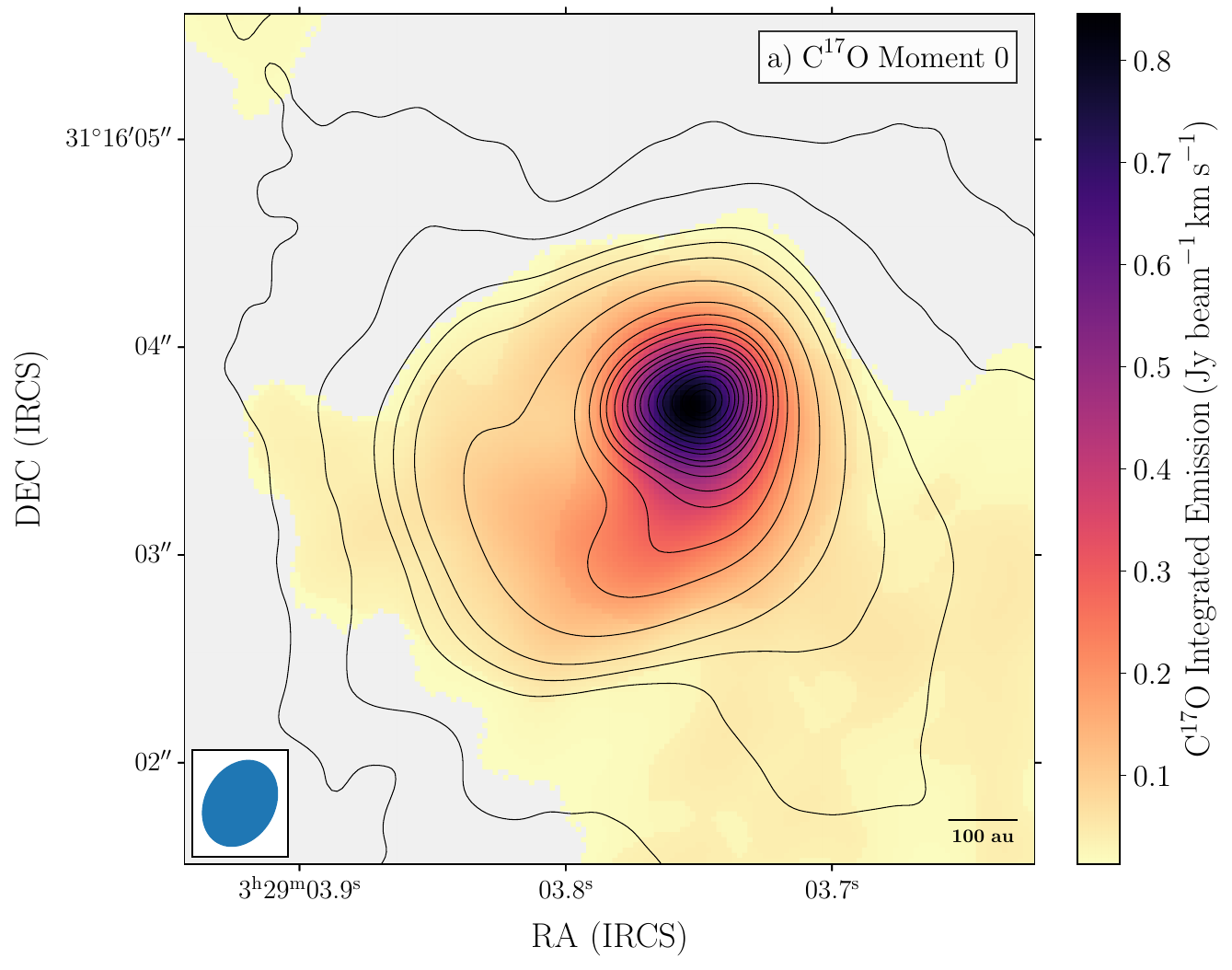}
\includegraphics[width=0.49\hsize]{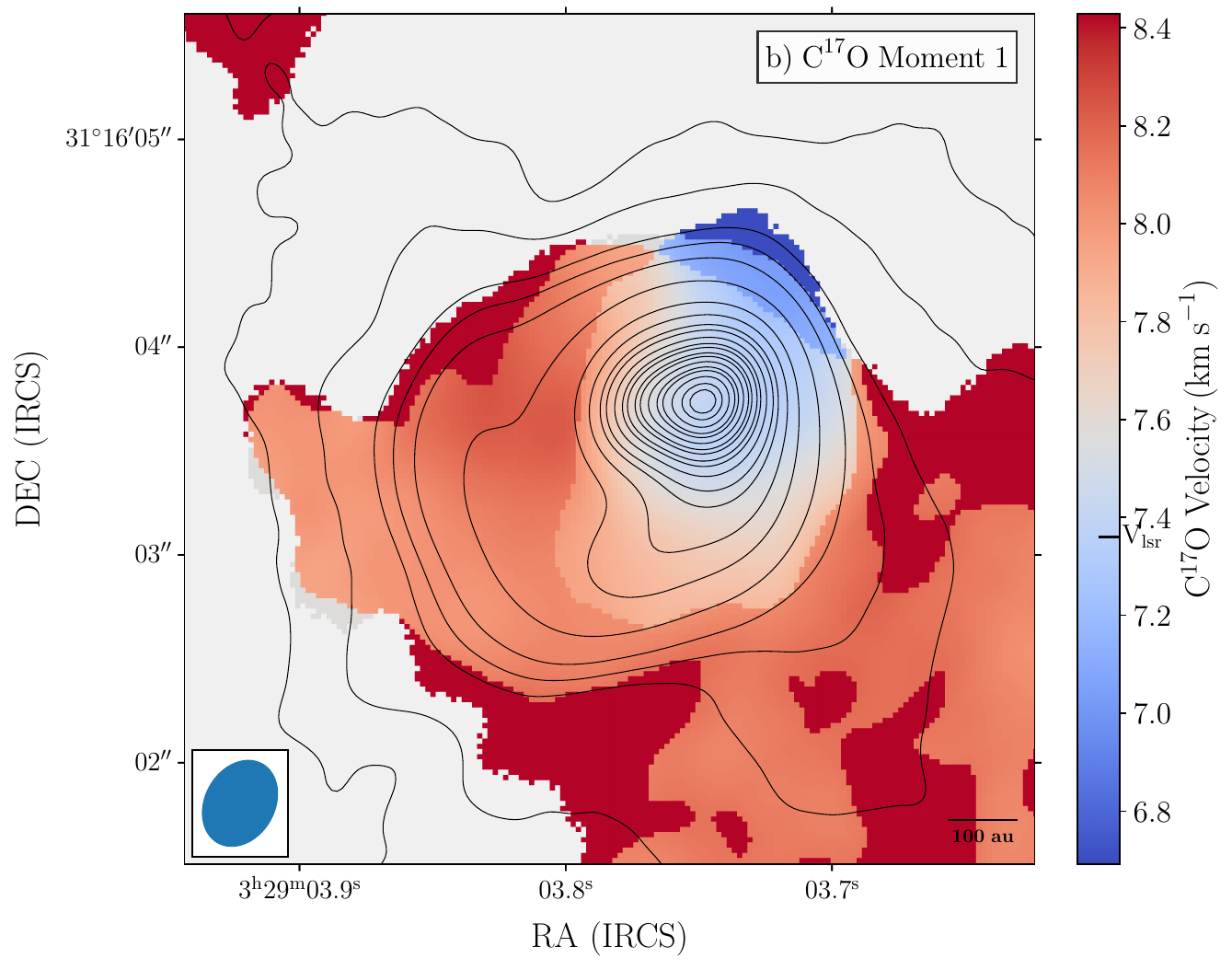}\\
\includegraphics[width=0.49\hsize]{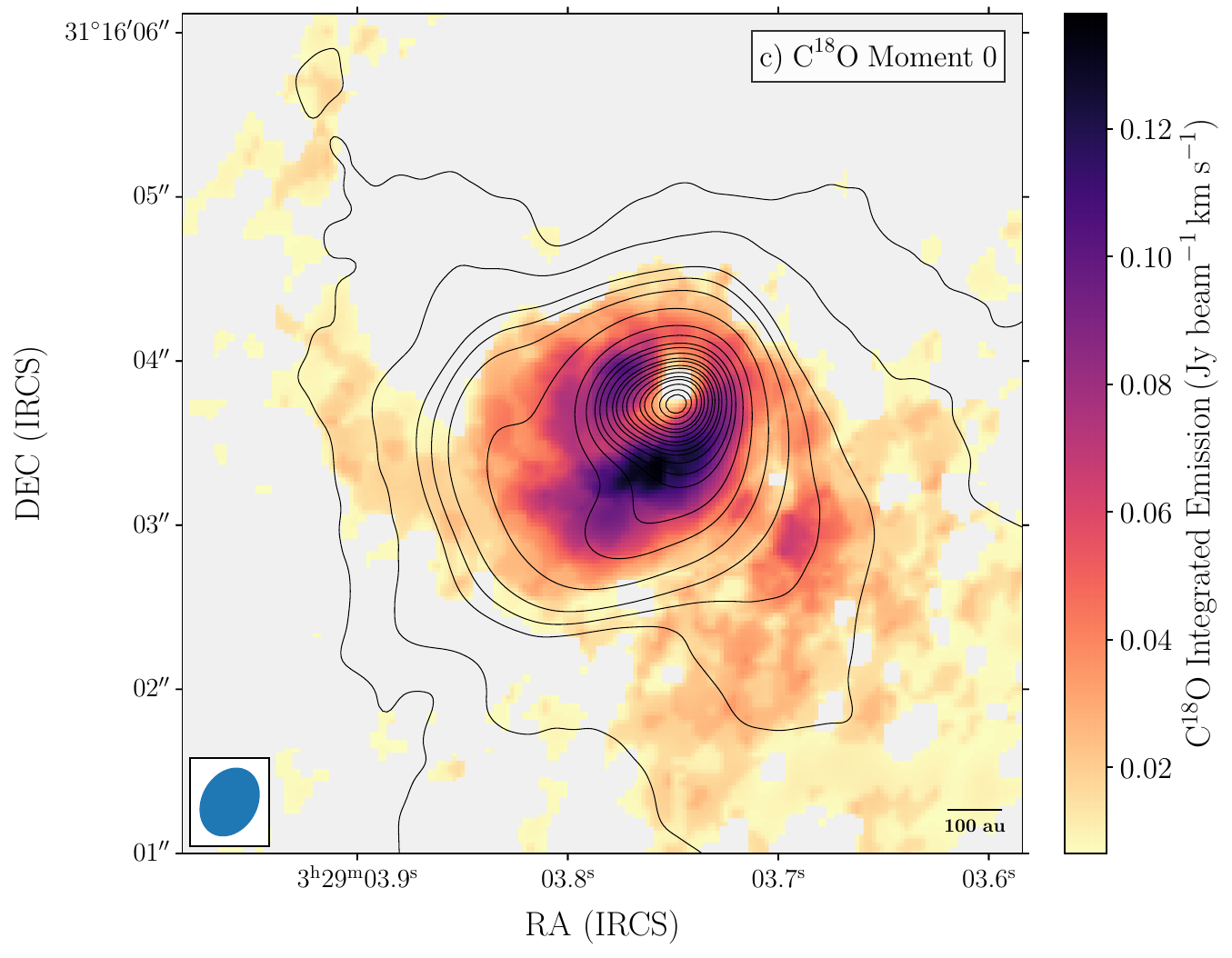}
\includegraphics[width=0.49\hsize]{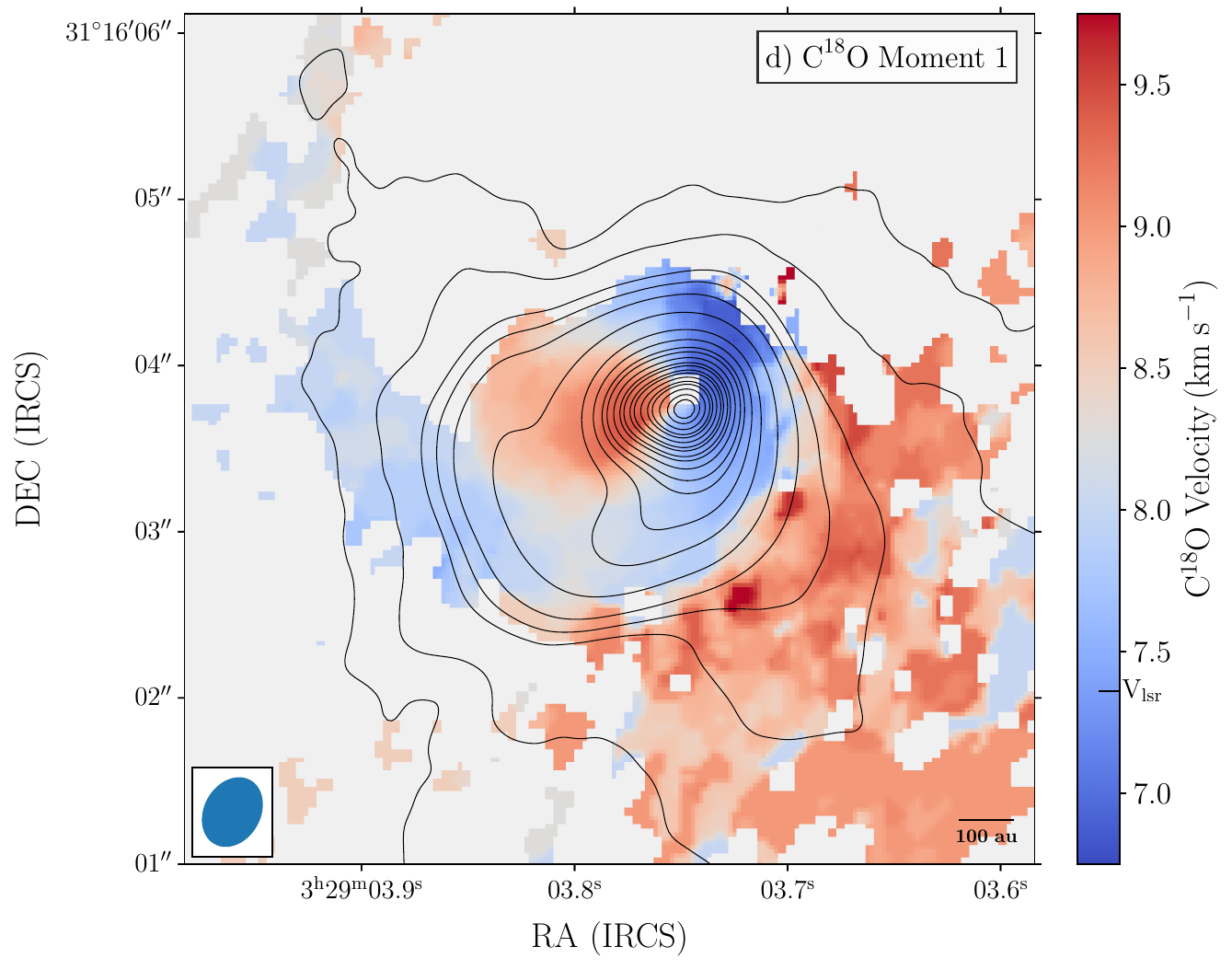}
\smallskip
\caption{{\bf a)} The C$^{17}$O$(J=3 \rightarrow 2)$ moment 0 map imaged using robust=1 is superposed with the contours from the dust continuum emission at levels of 1.5, 3, 7, 10, 15, 28.7, 54.9, 81.1, 110, 130, 160, 190, 210, 240, 260, 290, 320, and 340 mJy beam$^{-1}$. The moment maps were produced considering pixels between 15 mJy beam$^{-1}$ to the peak emission in C$^{17}$O and channels between 8.4 {\kms} and 9.3 {\kms}. {\bf b)} Same as panel a), but for the moment 1 velocity field map. The dust continuum contours are superposed using the same map and levels.  Note that the V$_{\mathrm{lsr}}= 7.36 \,\,{\kms}$ is indicated at the colorbar in each moment 1 map. {\bf c)} In this panel we show the moment 0 map from C$^{18}$O$(J=2 \rightarrow 1)$ emission. Superposed on the map, we show the same contours from dust emission as in panel a). The  C$^{18}$O moment maps were produced under the same threshold as the C$^{17}$O maps and within a velocity range of 6.75 {\kms} to 10 {\kms}. {\bf d)} Same as panel c) but for the C$^{18}$O moment 1 map.  The relative size of the beam is shown as a blue ellipse in the bottom left corner,  and the scale is shown in the bottom right corner at each panel. 
 \label{fig:c17o}
}
\bigskip
\end{figure*}

\begin{figure*} 
\centering
\includegraphics[width=0.99\hsize]{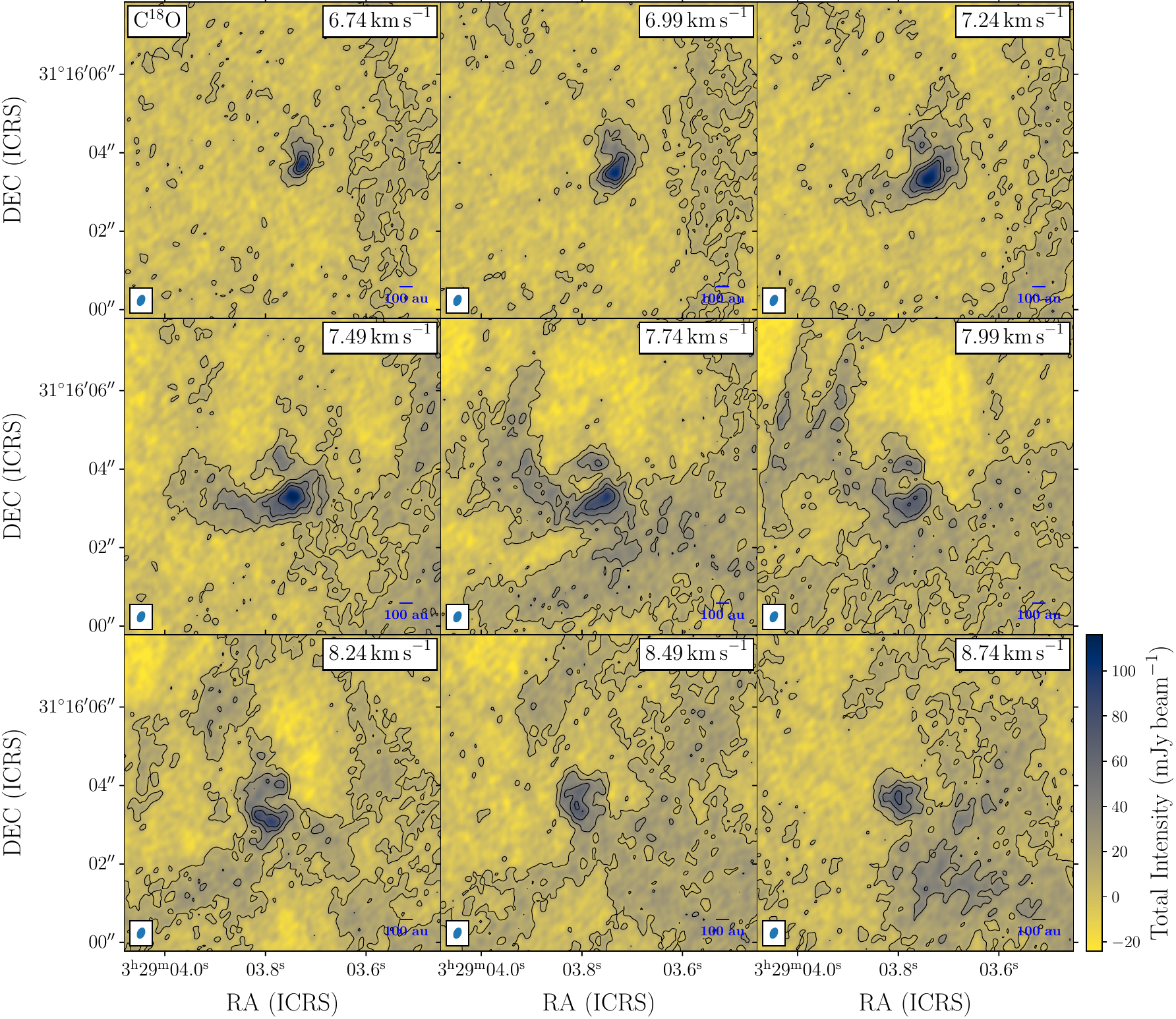}
\smallskip
\caption{Channel maps of C$^{18}$O emission shown in color scale between 6.7 {\kms} and 8.7 {\kms},  where the layout follows Figure \ref{fig:c17o_chan_map}. Superposed are C$^{18}$O contours in black with values 10, 30, 50, and 70 mJy beam$^{-1}$.
\label{fig:c18o_chan_map}
}
\bigskip
\end{figure*}

The C$^{18}$O$(J=2\rightarrow1)$ moment maps are shown in Figure \ref{fig:c17o}. The moment 0 map appears consistent with that of C$^{17}$O$(J=3\rightarrow2)$  where the spiral is now spatially resolved. 
Figure \ref{fig:c17o} also shows the moment 1 map, where the C$^{18}$O velocity field shows a gradient from 8.7 {\kms} to $\sim 7$ {\kms} along the spiral (the source $v_{lsr}$ is indicated in the colorbar). Close to the VLA 4 binary, there is evidence of rotation in the moment 1 map, as previously indicated. Despite the higher angular resolution of the data, there is still significant extended emission, particularly to the south of the circumbinary disk, which complicates the analysis. The presence of a wide-angle outflow in this source may introduce some kinematic contamination in the C$^{18}$O emission, which is difficult to untangle.  Nonetheless, the overall morphology of the spiral agrees quite well with the structures seen in the continuum and C$^{17}$O emission. Figure \ref{fig:c18o_chan_map} shows the C$^{18}$O channel maps where the spiral is resolved between 7.49 {\kms} and $\sim 8$ {\kms} which corresponds to the velocity range of a streamer previously detected in this source at larger scales from DCN$(J=3\rightarrow2)$ and also from C$^{18}$O$(J=2\rightarrow1)$ emission \citep{Hsieh2024}.

\subsubsection{\ce{DCN}\,$(J=3\rightarrow2)$}
\label{ssse:dcn}

\begin{figure*} 
\centering
\includegraphics[width=0.99\hsize]{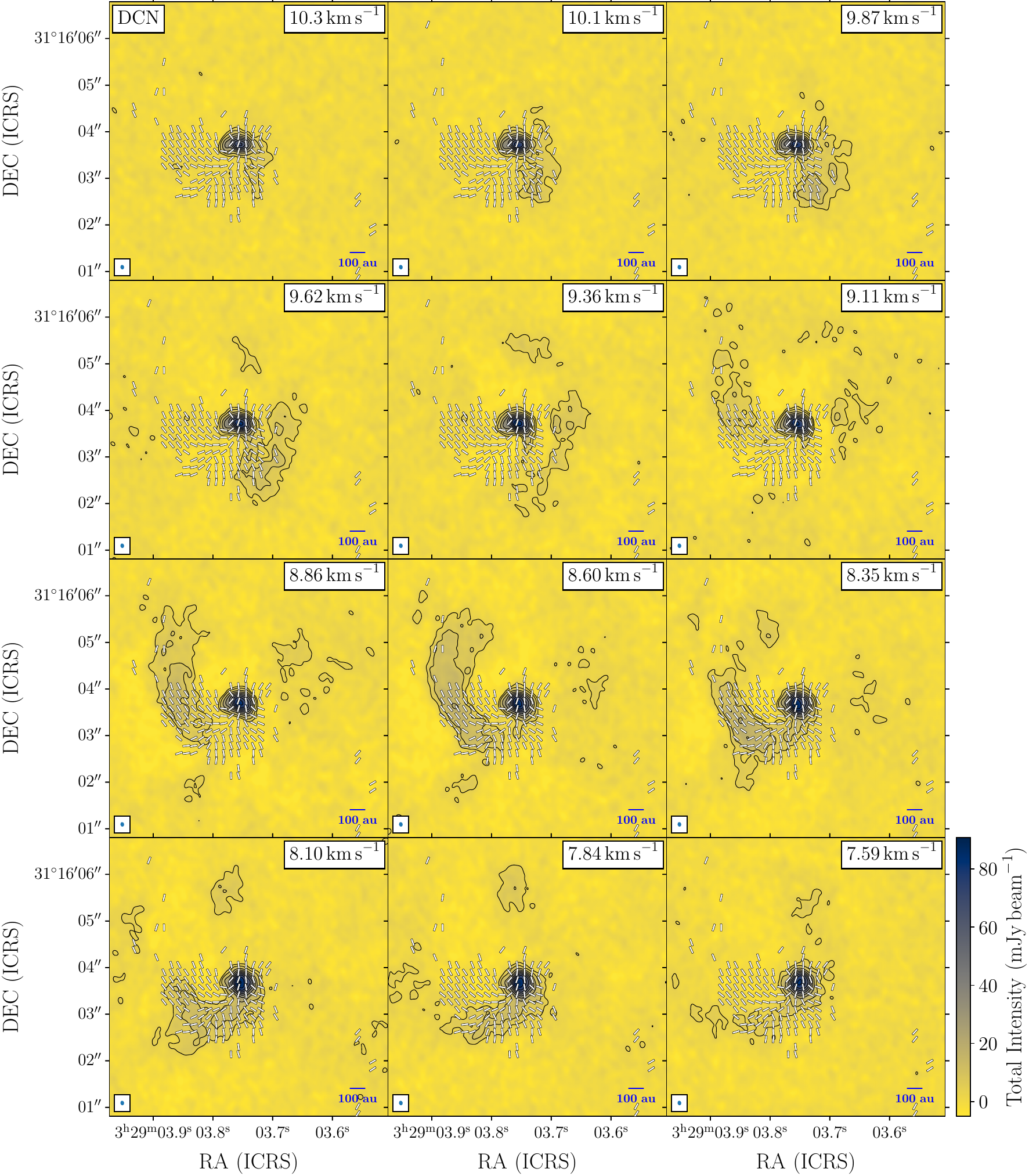}
\smallskip
\caption{Channel maps of DCN emission shown in color scale between  7.6 {\kms} to 10.3 {\kms}, also following Figure \ref{fig:c17o_chan_map}. Superposed are DCN emission contours in black with values of  4,   9,  19,  29,  39,  49,  59,  69,  80,  90, and 100 mJy beam$^{-1}$. At each channel, we also show the inferred magnetic field morphology from polarized dust emission as a map of white pseudo-vectors, plotted every 8 pixels to allow clearer comparison with the underlying DCN emission. The data were imaged using robust=2 and smoothed using a kernel of $0^{\prime\prime}.1$ to enhance the streamer.
\label{fig:dcn_chan_maps}
}
\bigskip
\end{figure*}

\begin{figure*} 
\centering
\includegraphics[width=0.99\hsize]{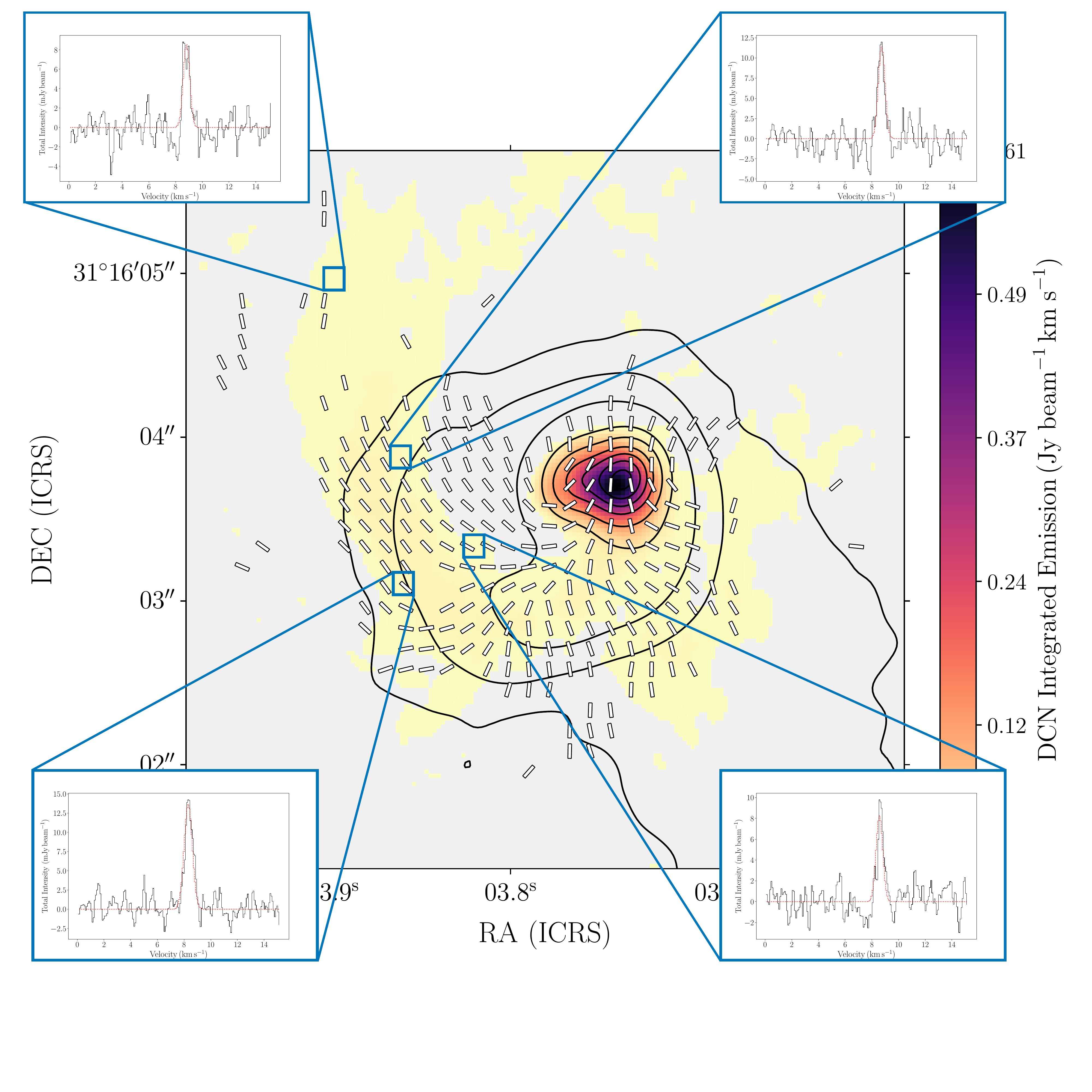}
\smallskip
\caption{DCN$(J=3 \rightarrow 2)$ moment 0 map integrated from 7.6 to 10.3 {\kms} is shown. Overlaid are dust continuum contours at levels of 3, 10, 28, 81, 130, 190, 240, and 290 mJy beam$^{-1}$ (robust = 0.5). The inferred magnetic field morphology is shown as white pseudo-vectors. DCN spectra from selected regions along the streamer are displayed in zoomed-in panels, with the corresponding Gaussian fits indicated by dotted red curves.
\label{fig:dcn_spectra}
}
\bigskip
\end{figure*}

As previously mentioned, \citet{Hsieh2024} mapped SVS13A with NOEMA and found evidence for a streamer in \ce{DCN}$(J=3 \rightarrow 2)$  and archival \ce{C^{18}O}$(J=2 \rightarrow 1)$ emission. Their results found a connection between the streamer and the tail of the dust spiral seen by \citet{Diaz-Rodriguez2022} and by extension with our results. As a follow-up to the NOEMA observations,
\ce{DCN}$(J=3\rightarrow2)$ was observed with ALMA toward SVS13A at a much higher angular resolution ($\sim 0^{\prime\prime}.1$) and spectral resolution ($\sim 0.1$ {\kms}). Emission from the DCN  molecule traces higher densities than C$^{17}$O and  C$^{18}$O \citep[$n_{\mathrm{crit}} = 6\times10^{6}$ cm$^{-3}$ for the $J=3\rightarrow2$ transition, ][]{Hsieh2023} and colder environments where carbon is expected to be frozen onto dust grains \citep{Caselli2002b}.
Figure \ref{fig:dcn_chan_maps} shows the DCN channel maps from 7.55 {\kms} to 9.07 {\kms}, which is within the range of the \citet{Hsieh2024} streamer and our  C$^{17}$O and C$^{18}$O channel maps. Over that velocity range, a spiral-like structure is indeed seen moving towards the circumbinary disk from the north-east and connecting with VLA 4 binary system. This spiral structure is likely the resolved NOEMA DCN streamer detected by \citet{Hsieh2024}. As with our other molecular line data, the DCN streamer matches the spiral seen in the dust emission extending up to $2^{\prime\prime}.25$ or  675 au, which is around the same length determined by \citet{Hsieh2024} from the NOEMA data. At velocities over 9 {\kms}, however, there appears to be another gas substructure coming from the West which also appears to connect with the circumbinary disk (see Figure \ref{fig:dcn_chan_maps}). This structure may be another streamer or gas coming from the bridge connecting to VLA3. This emission is faint and has too small an S/N to make any definitive conclusion. \citet{Hsieh2024}, fitted the NOEMA DCN hyperfine structure with a multi-Gaussian model, which confirmed optically thin emission. For the ALMA data, we fitted a simple Gaussian model to selected positions in the DCN velocity cube, obtaining FWHM velocity linewidths between 0.6 {\kms} and 0.7 {\kms}, which yields a mean velocity dispersion of 0.25 {\kms} (Figure \ref{fig:dcn_spectra} shows example spectra extracted from the data and their respective Gaussian fits). By using the streamer temperature estimate of $T_{\mathrm{s}} \sim 35-45$ K  \citep{Hsieh2024}, we obtained a streamer sound speed between 0.32--0.36 {\kms} ($\sim 0.34$ {\kms}), which suggests the streamer has subsonic motions.

\section{DISCUSSION}
\label{se:discussion}

\subsection{Interpreting the polarized dust emission}
\label{sse:interpreting}

Although the canonical interpretation for the origin of the polarized dust emission has been grain alignment by magnetic fields \citep{Crutcher2012},
observations of polarized dust emission from disks with ALMA, at similar scales, have revealed that other alignment mechanisms also can be in play. For instance, self-scattering explains the polarized emission seen in a number of proto-planetary disks \citep{Kataoka2016,Hull2018,Dent2019}, and even at coarser resolutions, ALMA has revealed evidence that self-scattering might also be detected, impacting the polarization position angle over the protostar \citep{Huang2024}. Although their detection of polarization in the $\beta$-Pic debris disk was marginal (2.7\,$\sigma$). \citet{Hull2022} found a result suggesting that radiative alignment torques and not magnetic alignment are the likely mechanisms behind the polarization detections in the disk. 
In addition, \citet{Looney2025} recently found a correlation between the position angles of dust filaments and the polarized dust emission, which they interpreted as the result of mechanical alignment.  In the case of SVS13A, we see a similar correlation close to the VLA 4 binary (see Figure \ref{fig:pol_map}, lower panel), which makes interpreting the polarization not straightforward. 
 Therefore, and at first glance, it is unclear if grain alignment by magnetic fields is the global mechanism behind the polarized emission from the dust in SVS13A.

To understand if the polarized dust emission obtained by ALMA corresponds to magnetically aligned dust grains, we compare the orientations of the total intensity emission and the C$^{18}$O velocity field gradients with respect to the inferred magnetic field orientation. Although the C$^{17}$O data share the same sampling of angular scales and were imaged using the same set of parameters (e.g., robust value), the line sampling in velocity space is not optimal because of the data coarse's velocity resolution ($\sim 2 \,\,{\kms}$ with respect to $\sim 0.2\,\, {\kms}$ of \, C$^{18}$O). We also compared these orientations to a model of the radiation field, which we disfavored and hence it is not discussed here (see Appendix \ref{se:appendix_a}). The comparison with the polarization position angle orientation is immediate as the field and the polarization position angle are orthogonal to each other. For this discussion, all quantities were computed using maps produced with robust=0.5, which gives the best balance between noise, angular resolution, and recovered emission.

\begin{figure*} 
\centering
\includegraphics[width=0.75\hsize]{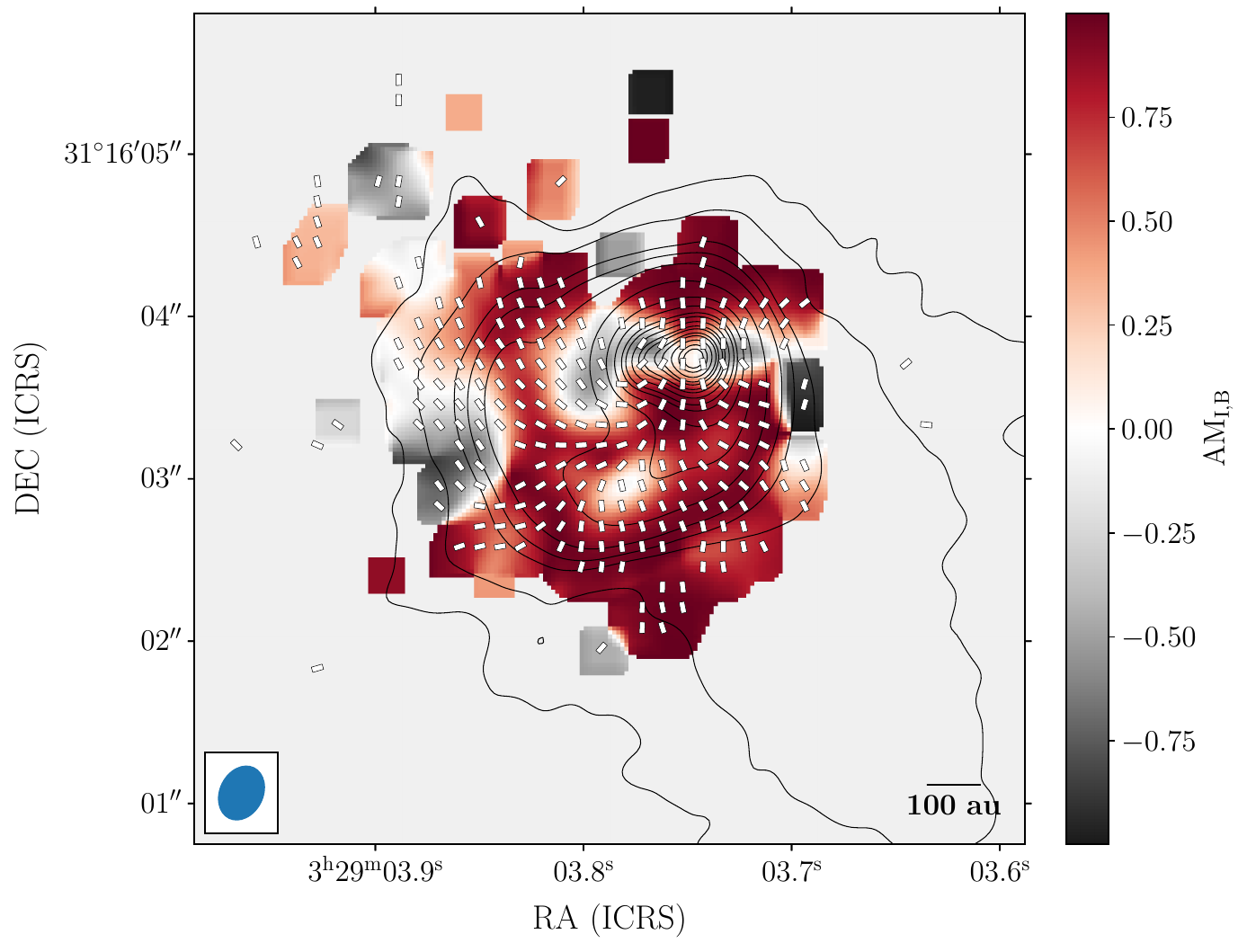}\\
\includegraphics[width=0.75\hsize]{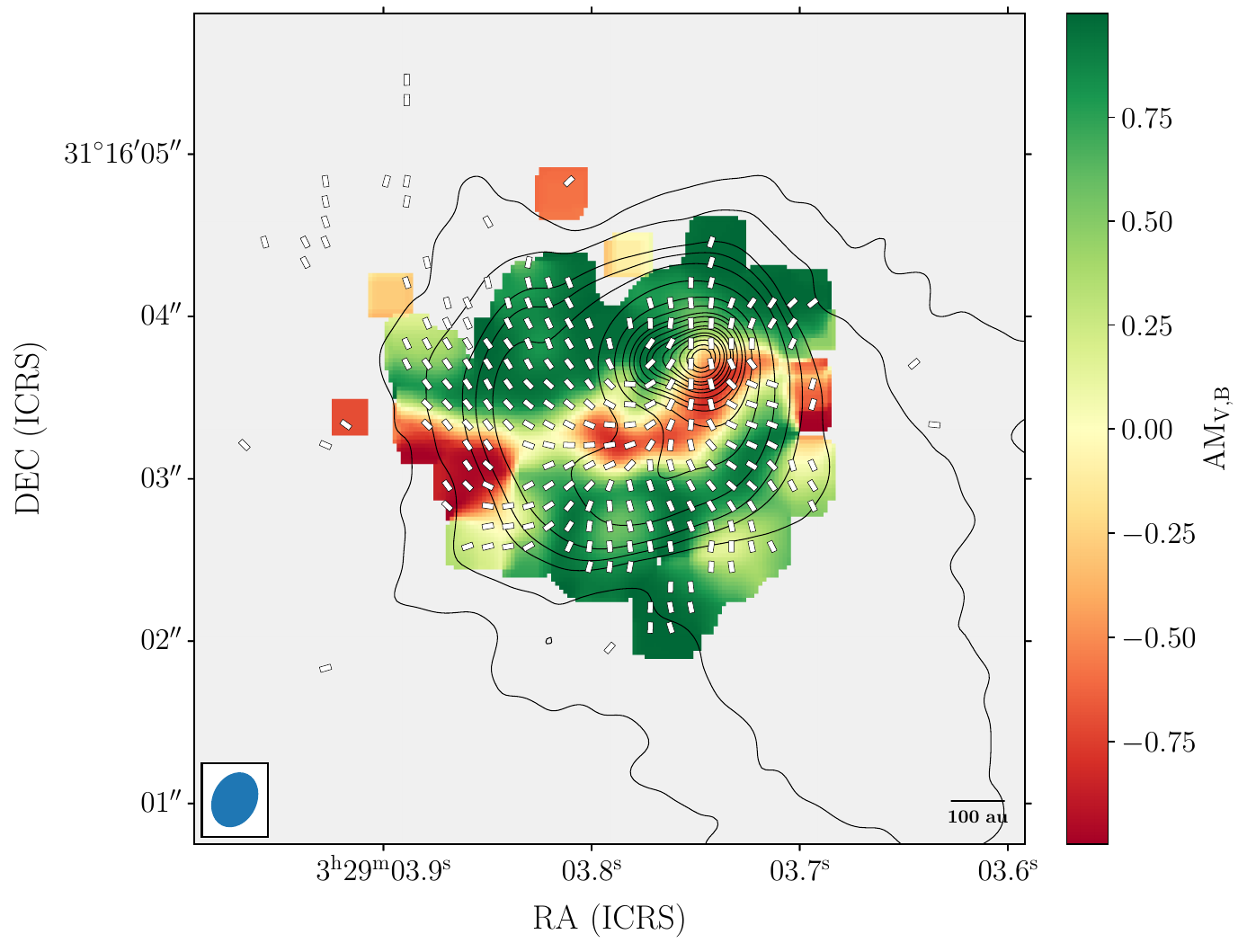}
\smallskip
\caption{{\bf\em Upper Panel.} In color scale, we show the AM map between the inferred magnetic field and the total intensity gradient orientation. Superposed is the magnetic field morphology onto the plane of the sky as white pseudo-vectors plotted every 5 pixels, which is approximately Nyquist sampling. In contours, we show the continuum emission with levels following Figure \ref{fig:c17o}.  {\bf\em  Lower Panel.} Same as the upper panel, but here we show the AM map between the field and the velocity gradient orientation. The size of the beam is shown as a blue ellipse in the bottom left corner,  and the scale is shown at the bottom right corner in each panel. 
\label{fig:am_map}
}
\end{figure*}

We compute the gradients following the approach implemented by \citet{Soler2013}, who used Gaussian derivatives (see Appendix \ref{se:appendix_a}). By visually inspecting the magnetic field morphology with the maps of the intensity and velocity gradients, we find similar orientations between them throughout most of the region, particularly to the north-east and to the south of the VLA 4 binary system  (see Figures in sections \ref{sse:int_gradient} and \ref{sse:v_gradient}). To quantify the comparison between the field, the intensity gradient, and the velocity gradient orientations, we compute maps and histograms of the alignment measure \citep[AM, ][]{Gonzales2017,Gonzales2019}, which is defined as:
\begin{equation}
\mathrm{AM} = \left<\cos{(2\Delta\theta_{\mathrm{1,2}}}) \right>~,
\end{equation}
where $\Delta\theta_{\mathrm{1,2}} = |\theta_{\mathrm{1}} - \theta_{\mathrm{2}}|$ represents the absolute angle difference between either the total intensity gradient and the magnetic field ($\Delta\theta_{\mathrm{I,B}}$), or the velocity gradient and the magnetic field ($\Delta\theta_{\mathrm{V,B}}$). Values for the AM parameter $\sim$ 1 indicate that the average angular difference in the directions of the two pseudo-vectors across the map is small; if AM $\sim$ 0, then there is no alignment between the two. But if AM $\sim -1$,  the two directions tend to be perpendicular to each other \citep{Lazarian2018}. We calculate the AM parameter pixel by pixel over the regions where there is emission overlap, but considering only the extension of the circumbinary disk. 
The average is obtained by taking a moving window (see section \ref{sse:b_streamer}), as we later do for the field dispersion following the approach used by \citet{Cortes2024}. 
\begin{figure*} 
\centering
\includegraphics[width=0.45\hsize]{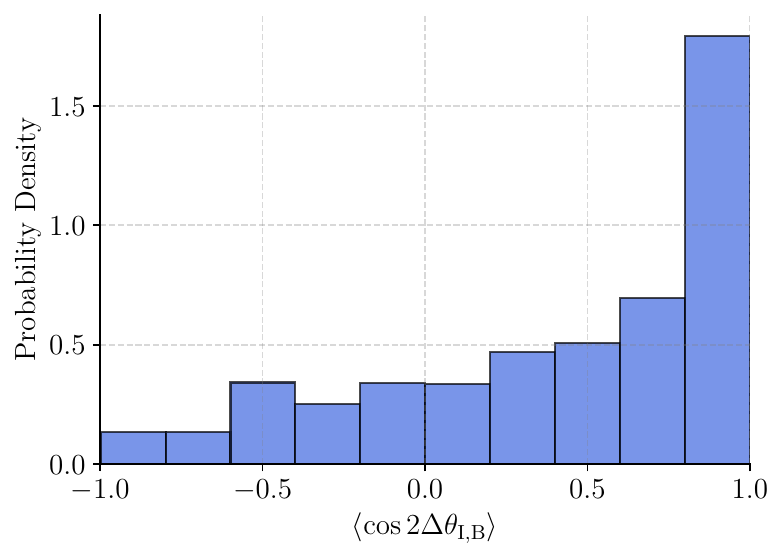}
\includegraphics[width=0.45\hsize]{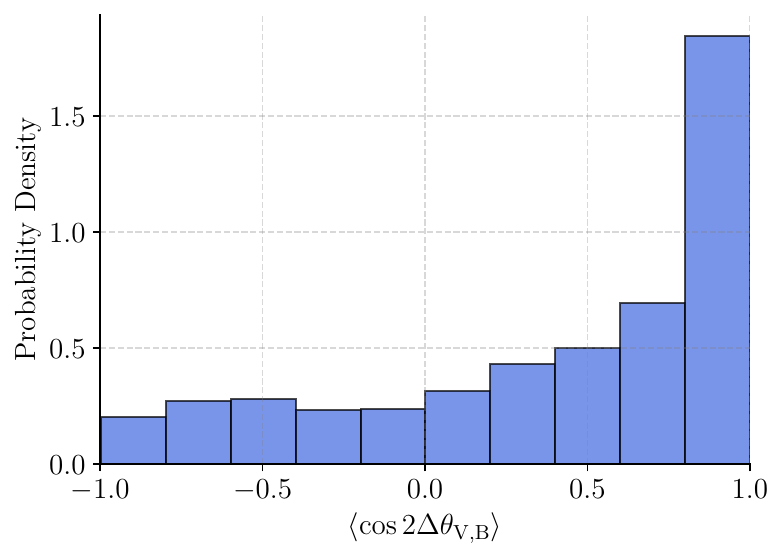}

\caption{{\bf\em Left panel.} The histogram of the AM parameter between the total intensity gradient orientation and the magnetic field is shown as a probability density. {\bf\em Right panel}. Same as the {\em left panel}, but here we show the AM parameter between the velocity gradient orientation and the magnetic field.
\label{fig:am_hist}
}
\bigskip
\end{figure*}
Figure \ref{fig:am_map} shows the AM maps. Visual inspection confirms agreement in the orientation to the north-east and south of the VLA 4 binary system between the gradients and the inferred magnetic field directions. 
Figure \ref{fig:am_hist} shows the histograms of AM values for the angle differences, where we see a larger probability around 1. 
Although the histograms also show a population of intermediate values, the preferred orientation is alignment between the inferred magnetic field orientation and the intensity and velocity gradients. 
We ignore the intermediate population as the associated probability is small. Alignment between the polarization position angle and the gradients shows a very low probability (AM $\sim -1$) and therefore is disfavored. 
The nature of these regions is not obvious, and we explore possible explanations in Section \ref{sse:b}.
Based on these results, the simplest conclusion is that we are seeing similar orientations between the intensity, velocity gradients, and the inferred (projected) magnetic field morphology over most of the SVS13A circumbinary disk.

\subsection{Alternative Grain Alignment Mechanisms}
\label{sse:grain_alignment}

\subsubsection{Self-Scattering}

 When dust grain sizes are on the order of the wavelength, the dust emission can self-scatter, producing a uniform polarization pattern oriented with respect to the minor axis of the disk. The expected fractional polarization values for self-scattering are $\leq 1\%$ \citep{Kataoka2016}. The fractional polarization over the VLA 4 binary (see Figure \ref{fig:pfrac}) indicates values in the self-scattering range. Furthermore, the higher resolution observations of \citet{Diaz-Rodriguez2022} put the VLA 4A and 4B disks with their minor axes slightly aligned to the polarization position angle that we see \citep[a direction orthogonal to the field; see Figure 8 in ][]{Diaz-Rodriguez2022}. Although we are not able to resolve the VLA 4A and B disks in our data,  one or two beams may be consistent with polarization from self-scattering. The inferred magnetic field over the central region of the map, however, is also consistent with a quite uniform, almost pinched, shape on scales larger than what we can expect from self-scattering coming from individual disks. Similar situations have been seen in other sources, particularly from the BOPS survey \citep{Huang2024}. Although self-scattering may be present in SVS13A, its extent is most likely confined to a few pixels over the VLA 4 binary.

 \subsubsection{Mechanical Alignment}
 \label{sse:mech}

 Because of the apparent correlation between the spiral structure observed in both the dust and gas emission moment 1 maps with the polarization position angles in the SVS13A circumbinary disk near the VLA 4 binary (see Figure \ref{fig:pol_map}, upper panel, and Figure \ref{fig:c17o}), mechanical alignment of dust grains emerges as a plausible explanation for the origin of the polarized emission. The underlying physics of this mechanism, initially proposed by \citet{Gold1952}, has since evolved but retains its fundamental principles \citep{Lazarian2007,Reissl2023,Lin2024}. In this scenario, a gas flow can align elongated dust grains along the flow direction, producing linearly polarized emission with position angles parallel to the flow. A comparison between the velocity gradients and the polarization pseudo-vectors, however, reveals that the flow direction is predominantly orthogonal to the polarization pattern, which is inconsistent with mechanical alignment. Nevertheless, there are regions—particularly to the east and south of the VLA 4 binary—where the polarization does appear to align with the local velocity gradients (see Figure \ref{fig:am_map}), which suggests that mechanical alignment may play a role locally.
 Recently and inspired by the Badminton Birdie, \citet{Lin2024} proposed a novel modification to the mechanism where the center of mass of the grain can be slightly off the geometric grain center. Under a steady flow, this asymmetry will produce a restoring torque that has the potential of aligning such a grain configuration to the gas flow, even if the flow is subsonic. The calculation assumes the flow to be under the Epstein regime \citep{Epstein1924}, where the grain size is smaller than the mean free path, which reduces the problem to an elegant damped harmonic oscillator equation. The proposed process is efficient, aligning a grain in $\sim 4$ days at $\sim 1$ au and in $\sim 10^{3}$ years at $\sim 100$ au from the star, based on a protoplanetary disk model. 
 
 In contrast, the magnetic alignment time scale depends on the radiation field to supra-thermally spin-up the grain to trigger a Larmor precession to align the grain's angular momentum with the field direction. The former depends on the strength of the radiation field while the latter depends on the strength of the magnetic field \citep[see equation 87 in ][]{Lazarian2007}. The strength of the radiation field in SVS13A is difficult to estimate, but it has been established that both VLA 4 sources have evidence of hot-corino chemistry whereas the gas around the binary system and in the circumbinary disk has temperatures in excess of 100 K. Because of this, we can assume a strong radiation field which suggests efficient supra-thermal spin-up of grains by the RAT mechanism is possible. Thus,  using a $n_{\mathrm{H_{2}}} \sim 10^{9}$ cm$^{-3}$ estimate for the circumbinary disk \citep{Diaz-Rodriguez2022} and a conservative assumption for the field strength of 500 $\mu$G obtained by extrapolating from Zeeman values \citep{Crutcher2019}, the Larmor precession time-scale is $\sim 10$ hours. Note that this field strength value is smaller than our estimate in the streamer (see Section \ref{se:discussion}). This suggests that overall alignment by magnetic fields is faster than the Badminton-Birdie mechanical alignment in SVS13A. If the radiation field is attenuated and cannot efficiently spin up the grains, however, the Badminton-Birdie mechanism might be faster, and the polarized emission seen close to the VLA 4 binary might be the result of mechanical alignment. If true, this could be a potentially exciting new research window because it offers the possibility of mapping dust kinematics through linearly polarized dust emission. The applicability of this mechanism should be further explored.
 
\subsection{Are we seeing magnetic fields?}
\label{sse:b}

The results obtained from maps and the histograms of the AM parameter favor similar orientations between the gradients and the magnetic field direction throughout most of the SVS13A circumbinary disk. The orthogonal case, alignment with the polarization position angle, is not favored by the results. We interpret this result as evidence for magnetically aligned dust grains in the SVS13A circumbinary disk. Although other mechanisms, such as self-scattering and mechanical alignment, may be present, these may be acting only at locations that are not directly related to the 
streamer. Furthermore, the SVS13A region is an evolved class I binary system with a relatively large disk that exhibits a clear spiral structure. In evolved sources, where stars have already formed, it is expected that the magnetic field will no longer oppose gravitational forces effectively close to the stars. Thus, the simplest explanation is that the field might have the same orientation as the gravitational field, where its proxy is the intensity gradient. Additionally, alignment with the velocity field is what we would expect if the field is channeling the gas as it infalls onto the circumbinary disk. This alignment occurs because the gas will tend to move more freely parallel to the field lines than across them. Hence, this polarized emission may be the first evidence for a magnetically dominated streamer, as we will argue in section \ref{sse:b_streamer}.

Although our results suggest magnetic alignment of dust grains throughout most of the SVS13A region, there are exceptions around the VLA 4 binary system and to the East of the circumbinary disk, where there is almost no alignment with either the field-intensity and field-velocity AM maps (see Figure \ref{fig:am_map}). Because self-scattering is likely constrained just to a few pixels over the binaries, one possibility is that at those locations the total intensity may be contaminated by free-free emission as suggested by the $0^{\prime\prime}.5$, 1.3 and 3 cm continuum emission coming from the binaries \citep{Diaz-Rodriguez2022}. Since free-free emission is not linearly polarized, however, the polarized continuum emission that we detect over there is likely coming from dust within an envelope. An envelope would alter the correlation between the intensity gradient and the polarization direction as suggested by the lack of polarized intensity around VLA 4A (see Figure \ref{fig:pol_map}). The lack of alignment to the East, however, is more puzzling because there is no indication of free-free contamination in the continuum at that position and the C$^{18}$O emission appears to follow the dust spiral (see Figure \ref{fig:c17o}). The channel maps do reveal some extended  C$^{18}$O emission to the East between 8 {\kms} and 8.24 {\kms}, which may not be associated with the spiral/streamer, but such emission may be locally contaminating the velocity gradient map. Otherwise, the origin of this discrepancy remains unclear.

\subsection{A sub-Alfv\'enic Streamer } 
\label{sse:b_streamer}

\begin{figure*} 
\centering
\includegraphics[width=0.95\hsize]{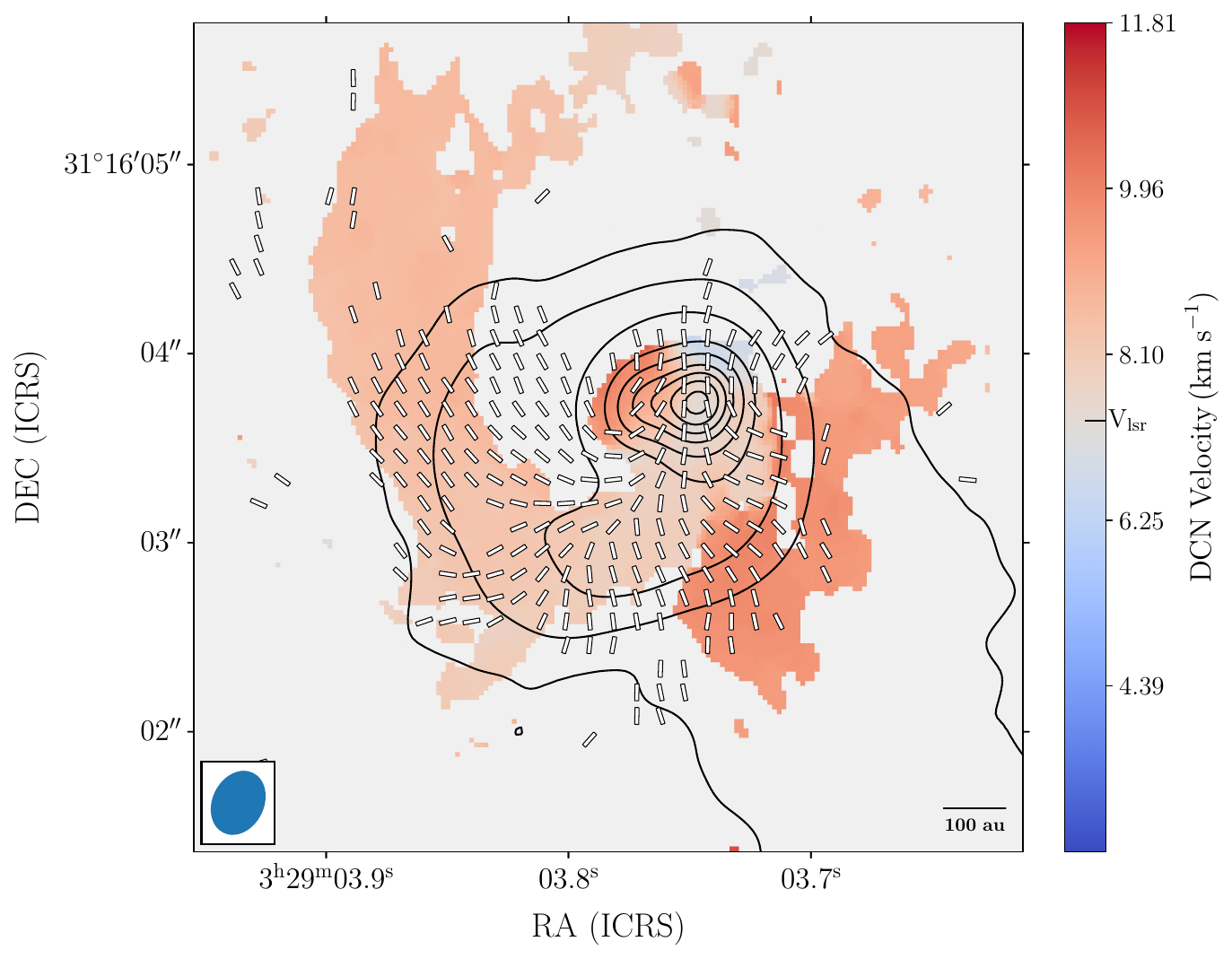}
\smallskip
\caption{The moment 1 map from DCN emission using the same layout as Figure \ref{fig:dcn_spectra} with the SVS13A systemic velocity indicated in colorbar (V$_{\mathrm{lsr}} = 7.36\,\, {\kms}$). The beam is depicted as a blue ellipse in the bottom left corner, and the scale is displayed in the bottom right corner.
\label{fig:dcn_field}
}
\bigskip
\end{figure*}

As with C$^{17}$O and C$^{18}$O, Figure \ref{fig:dcn_field} shows how the DCN moment maps correlate with the magnetic field. The DCN velocity channels are imaged against the magnetic field in Figure \ref{fig:dcn_chan_maps}. From the Figures, it is clear that there is a spatial correlation between the streamer and the field, where the subsonic streamer follows the magnetic field orientation from the north-east up to the circumbinary disk (see section \ref{ssse:dcn}). Close to the protostars, the agreement between the field and the streamer is less clear because the magnetic field changes orientation with respect to the streamer morphology. As previously mentioned, however, it appears that a second gas substructure is present south-west of the binary system but at slightly higher velocities (from $\sim$ 9.1 {\kms} to 10 {\kms}, also see  Figure \ref{fig:dcn_chan_maps}), but with less significance. 
This second structure appears to align with the field towards the south-west of the disk, closer to the binary system. Because of the low S/N of this structure, and because we cannot definitively establish grain alignment by magnetic fields in that part of the circumbinary disk, we will refrain from interpretation here, leaving further investigation for future work. The following analysis only refers to the north-east section of the streamer, over the bifurcation line, where we have established grain alignment by magnetic fields.

To understand if the gas in the streamer preferentially flows along the field lines, where the field provides pressure in the orthogonal direction, or if the gas and the field are just coupled without the latter providing meaningful dynamical support, we need to compare the turbulent energy against the magnetic energy, as the system is already assumed to be gravitationally dominated.
To this effect, estimating the magnetic field strength is required, which we do by applying the Davis-Chandrasekhar-Fermi method \citep[or DCF,][]{Davis1951,Chandrasekhar1953}, or:
\begin{equation}
\label{eq:dcf}
\mathrm{B}_{\mathrm{pos}}=\sqrt{4\pi\rho}\,\sigma_{v}/\delta_{\phi}~,
\end{equation}
where $\rho$ is the volume density, $\sigma_{v}$ is the velocity dispersion, and $\delta_{\phi}$ is the dispersion in the magnetic field lines. In practice, the magnetic field strength onto the plane of the sky can be expressed as $\mathrm{B}_{\mathrm{pos}} = 9.3 \sqrt{n_{\mathrm{H}_{2}}} \Delta V / \delta \phi$\, $\mu$G, where $n_{\mathrm{H}_{2}}$ is the number density, and $\Delta V$ is the FWHM linewidth obtained from a Gaussian fit to the line data \citep{Crutcher2004}. Although estimating the number density requires information about the temperature and the source geometry, in SVS13A we can use global parameters to obtain an estimate for $\mathrm{B}_{\mathrm{pos}}$.  Assuming the critical density from DCN$(J=3 \rightarrow 2)$, or $n_{crit} \sim 6\times10^{6}$ cm$^{-3}$, as the gas number density when considering $T \sim 40$ K for the streamer \citep{Hsieh2023}, a mean velocity linewidth, $\langle \Delta V \rangle = 0.6 \,{\kms}$ (see Section \ref{ssse:dcn} and Figure \ref{fig:dcn_spectra}), and a field dispersion over the streamer of $\delta_{\phi} = 12^{\circ}$ (calculated as the circular standard deviation), we obtain $\mathrm{B}_{\mathrm{pos}} = 1.1 \pm 0.6$ mG,  where the uncertainty is calculated using error propagation over equation \ref{eq:dcf}. The uncertainty in the critical number density is likely the largest factor because of the lack of collisional rates for DCN \citep{Hsieh2024}. This absence can have an effect on the uncertainty as large as a factor of 3 \citep{Shirley2015}, which we use here. We used 0.1 {\kms} in the linewidth (coming from the velocity resolution), and a polarization position angle error of $2^{\circ}$ for the ALMA observations (see Appendix \ref{sse:small_window}).

To estimate the magnetic energy, we evaluate
\begin{eqnarray}
E_{B} &=& \frac{1}{8\pi}\int B^{2}d^{3}x \nonumber \\
      &=& \frac{1}{8\pi}\mathrm{B}^{2}_{\mathrm{pos}} \times \mathrm{V}~,
\end{eqnarray}
where $\mathrm{V}$ is the volume assuming a constant field. The kinetic energy in the streamer is similarly estimated as
\begin{equation}
E_{k} = \frac{1}{2}\rho v^2 \times \mathrm{V}~,
\end{equation}
where $v$ is the streamer's gas velocity, and $\rho = m_{\mathrm{H}_{2}}n_{\mathrm{H}_{2}}$ is the volume density, with $n_{\mathrm{H}_{2}}$ and $m_{\mathrm{H}_{2}}$ the number density and molecular hydrogen mass respectively. To compare both energies, we take the ratio, which cancels the volume, yielding
\begin{equation}
\label{eq:en_ratio}
\frac{E_{k}}{E_{B}} = \frac{4\pi m_{\mathrm{H}_{2}}n_{\mathrm{H}_{2}} v^2}{B_{\mathrm{pos}}^{2} }~.
\end{equation}

If we consider $v^{2}=2\sigma_{u}^{2} + c_{s}^{2}$ to include both thermal and non-thermal motions \citep{Myers1988}, where $\sigma_{u}=\left< \Delta V \right>/2\sqrt{2\log{2}}=0.22 \,{\kms}$ is the velocity dispersion along the line-of-sight after removing the channel width, with the factor of 2 representing the assumption that the velocity on the plane is the same as the line-of-sight, and $c_{s} = 0.34\, {\kms}$ is the sound speed for $T_{gas}$=40 K. We obtain $E_{k}/E_{B}=0.5 \pm 0.4$, which suggests that magnetic energy regulates the gas flow in the subsonic streamer.
As with $B_{\mathrm{pos}}$, the uncertainty in the energy ratio was computed using error propagation over equation \ref{eq:en_ratio}. It should be noted that the estimated uncertainty suggests that the regulating effect of the magnetic field in the infalling flow appears to be marginal, as both quantities are not far from equipartition. This estimate yields a single value that should be regarded as an average representation rather than a detailed description of local conditions.

From Equation \ref{eq:en_ratio}, it can be shown that
\begin{equation}
\frac{E_{k}}{E_{B}} = \frac{1}{2}\mathcal{M}_{A}^{2}
\end{equation}
where $\mathcal{M}_{\mathrm{A}} = \frac{\Delta V}{V_{\mathrm{A}}}$ is the Alfv\'en Mach number, with $V_{\mathrm{A}}=\mathrm{B}_{\mathrm{pos}}/\sqrt{4\pi\rho}$ an estimate of the Alfv\'en speed. Thus, the $\mathcal{M}_{\mathrm{A}}$ number can also be understood as a representation of the ratio between kinetic and magnetic energy and therefore can also indicate the relative importance between the magnetic and the gas kinetic energies. 
Furthermore, \citet{Cortes2024} showed that when assuming DCF, the $\mathcal{M}_{\mathrm{A}}$ number can be directly estimated from the field dispersion, yielding a map. Namely, 
\begin{eqnarray}
\mathcal{M}_{\mathrm{A}}
  &=& \frac{\Delta V}{V_{\mathrm{A}}} \nonumber\\
  &=& \frac{\Delta V}{B/\sqrt{4\pi\rho}} \nonumber\\
  &=& \frac{2\sqrt{2\log 2}\,\pi}{4\,Q}\,\delta\phi
\end{eqnarray}
where $\Delta V = 2\sqrt{2\log{2}}\,\sigma_{v}$ is the FWHM measure of the line, $B=\frac{4}{\pi}\,Q\,\mathrm{B}_{\mathrm{pos}}$ where $Q=0.5$ is a correction factor derived from simulations \citep{Crutcher2004, Liu2022} and $\frac{4}{\pi}$ corresponds to the statistical average of the magnetic field when considering isotropic 3-dimensional turbulence \citep{Crutcher2004} and $\delta \phi$ is the local dispersion of the magnetic field lines. 
The $\mathcal{M}_{\mathrm{A}}$ derived here, is analogous to the ``local'' Alfv\'en Mach number defined by \citet{Burkhart2009}.
The difference is that in their work, they considered values obtained from Zeeman measurements, while we use DCF estimates.
It should be noted that the estimate of $\mathcal{M}_{\mathrm{A}}$ derived here is as good as DCF itself.

\begin{figure*} 
\centering
\includegraphics[width=0.99\hsize]{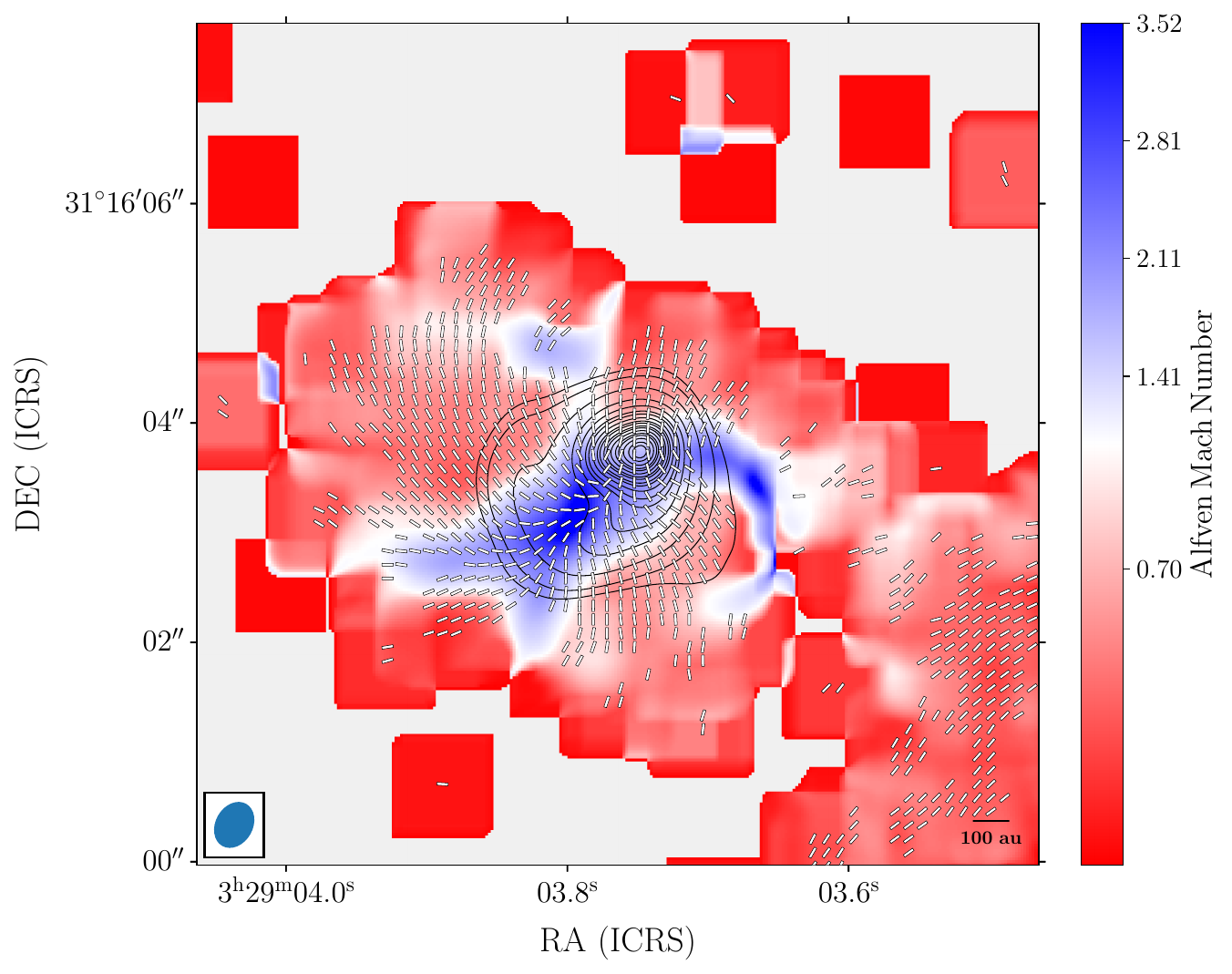}

\caption{The Alfv\'en Mach number from the SVS13A magnetic field map is shown in a color scale. The color white represents the trans-Alfv\'enic value of $\mathcal{M}_{\mathrm{A}} = 1$ while red indicates the sub-Alfv\'enic regime.
The magnetic field morphology is superposed as white pseudo-vectors plotted every five pixels.
The map was produced from the data imaged using robust=1. The beam is depicted as a blue ellipse in the bottom left corner, and the scale is displayed in the bottom right corner. 
\label{fig:Alfven_mach_number}
}

\end{figure*}

To obtain a map of $\delta_{\phi}$, we followed \citet{Cortes2024}, who applied a moving window to the polarization position angle data. Assuming that the distribution of the polarization position angle is Gaussian, one would obtain a statistical error of $\sim 6^{\circ}$ when using a $1^{\prime\prime}$ window, or about 9 independent points, for our SVS13A field map \citep[see ][ Appendix B for a derivation of the error and Appendix \ref{sse:small_window} in this work]{Cortes2024}. Figure \ref{fig:Alfven_mach_number} shows the $\mathcal{M}_{\mathrm{A}}$ map when using a moving window of $1^{\prime\prime}$ and a robust value of 1 to recover sufficient emission over the streamer. The resulting map suggests that the gas motions along significant portions of the streamer are sub-Alfv\'enic ($< 1$), or that the magnetic energy is larger. In practice, this means that magnetic field tension is constraining movements orthogonal to the field lines as the neutral gas is coupled to the field through collisions with the charge carriers. Thus, the preferred direction of motion is along the field lines, which sets the morphology of the infalling flow. As explained before, $\mathcal{M}_{\mathrm{A}}$ is a representation of the energy balance between kinetic (non-thermal) and magnetic energy.
This result is in line with the subsonic gas motions determined in the streamer, which suggests that in the infalling material, turbulence is being dissipated by the magnetic energy, likely producing a laminar flow. 
The super-Alfv\'enic values seen close to the VLA 4 binary are likely the result of applying the moving window to the bifurcation seen in the field morphology (see Figure \ref{fig:pol_map}, lower panel). 
This effect becomes clear when examining an $\mathcal{M}_{A}$ map generated with a smaller moving window, though this also increases the statistical error (see Appendix \ref{sse:small_window}, Figure \ref{fig:am_map_05}).
We also note that the south-west substructure appears to be sub-Alfv\'enic as well, but we cannot conclude about its nature or its infalling properties, if any.

In principle, estimating $\mathcal{M}_{\mathrm{A}}$ directly from the dispersion in the field lines requires that the dispersion represent local deviations from the mean field. In general, this is challenging because we only have information on the position angle values projected onto the plane of the sky, while the 3-dimensional field structure is unknown. If the mean field is not subtracted, the estimated dispersion might be higher than what local perturbations by the gas might induce. This is because the deviations from the mean can be due to the large-scale field structure and not from local perturbations to the field lines. 
Although it would have been preferable to use as small a moving window as possible to estimate only the local field dispersion, there is a tradeoff between a larger moving window, which reduces the statistical error but propagates large angle dispersions from smaller to larger regions, and a smaller moving window which limits the propagation of larger dispersion angles, but has a small number of independent points increasing the statistical error.  Using a $1^{\prime\prime}$ moving window yields an error estimate of $\sim 6^{\circ}$, which is reasonable compared with the Band~7 polarization position angle error of $2^{\circ}$ (see Appendix \ref{sse:small_window}). In addition, the artificially larger dispersion values are confined to the center of the circumbinary disk, away from the streamer. 
The caveat of this approach comes from using the projected field morphology to compute a single dispersion map. If the field has multiple, uncorrelated, components along the line of sight, the $\mathcal{M}_{\mathrm{A}}$ map will not accurately represent the global Alfv\'en Mach number. In SVS13A, this can be the case at the locations where the field bifurcates. It is not possible, however, to obtain the 3-dimensional field morphology from the projected map without detailed physical modeling \citep[see ][]{Sanhueza2021,Saha2024}, which is outside the scope of this work.

\subsubsection{On the origins of the sub-Alfv\'enic streamer in SVS13A}

Streamers are thought to channel material from the envelope onto the disk, thereby feeding the growth of both disk and protostar \citep{Pineda2020}. In SVS13A, the streamer reported with NOEMA and now resolved by our ALMA data fits into this picture, extending over $\sim$700 au \citep{Hsieh2023}. Although our observations cannot directly trace the connection between the circumbinary disk and the larger-scale envelope, the continuity of the magnetic field suggests that such a link must exist. This behavior makes the sub-Alfv\'enic character of the streamer particularly relevant as the magnetic field appears to facilitate a laminar infall from envelope scales onto the disk. 

At larger scales, JCMT/POL-2 polarimetry indicates a smooth and ordered magnetic field across the SVS13 region \citep[see Figure 4 in ][]{Doi2020}, while ammonia kinematics point to predominantly transonic to supersonic motions \citep[see Figure 9 in ][]{Dhabal2019}, suggesting moderate levels of turbulence in the envelope. This situation is consistent with other parts of NGC~1333, where velocity gradients suggest that colliding flows are seen throughout the cloud \citep{Dhabal2018,Dhabal2019}. From these data, it is unclear what physical parameter dominates or whether turbulence and magnetic energy are in equipartition as seen in several star-forming regions \citep{Pattle2023}. If the envelope were to be in equipartition ($\mathcal{M}_{\rm A}\!\sim\!1$), the formation of a sub-Alfv\'enic streamer would likely require localized conditions tipping the balance slightly toward the magnetic field. Such conditions may arise in regions that are locally more gravitationally supercritical, where gravity would bend the field channeling material inward along its lines. Alternatively, magnetic diffusion  
may facilitate slippage of neutral gas across field lines, allowing inflow along magnetically guided channels. 

Recent simulations by \citet{Tu2024} introduce ``gravo-magneto sheetlets'' as anisotropic inflow structures that bridge the envelope and disk, resembling observed streamers. Their model suggests a magnetically dominated envelope with gravity overwhelming the field at some points, bending the field lines due to ambipolar diffusion and allowing the flow to infall along the ``sheetlets'' feeding a disk. 
While our measured plasma-$\beta=0.7$\footnote{The plasma-$\beta$ is defined as the ratio of gas to magnetic pressure \citep{Choudhuri1998}} differs from the values obtained in the \citet{Tu2024} models, and although it remains uncertain whether the envelope in SVS13 is magnetically dominated, the \citet{Tu2024} framework nevertheless highlights how magnetic diffusion and the gravitational bending of field lines can naturally produce such channels. 
It should be noted, however, that filamentary, streamer-like flows are also found in a variety of numerical works that include both turbulence and magnetic fields \citep[e.g.,][]{Santos-Lima2012,Seifried2013,Li2014b,Seifried2015,Kuffmeier2017,Matsumoto2015,Gray2018,Lam2019,Wurster2020b,Mignon-Risse2021}. Investigating how sub-Alfv\'enic streamers can form from envelopes in equipartition should be pursued by numerical experiments. 

In summary, our results suggest that the streamer in SVS13A is magnetically sub-Alfv\'enic, bridging the envelope and disk through a spiral field configuration. The region where it meets the disk may mark the scale at which turbulence is injected into the circumbinary system.

\section{Summary and Conclusions}
\label{se:conclusion}

In this paper, we have presented the first results from the ALMA Perseus Polarization Survey (ALPPS),  a sub-Alfv\'enic streamer in SVS13A. Our ALMA observations include polarized and total intensity continuum emission along with C$^{17}$O$(J=3\rightarrow2)$, C$^{18}$O$(J=2\rightarrow1)$, and DCN$(J=3\rightarrow2)$ molecular line emission from the SVS13A circumbinary disk. The results can be summarized as follows,

\begin{itemize}

\item We report the first detection of a sub-Alfv\'enic infalling streamer. By comparing the molecular line emission morphology with the magnetic field map, we find a remarkable correlation that suggests the gas is coupled to the field. Although the system is likely gravitationally bound, the field appears to guide the streamer from the north-east (envelope) onto the SVS13A circumbinary disk. This interpretation is supported by our magnetic field strength estimate, $\mathrm{B}_{\mathrm{pos}} = 1.1 \pm 0.6$ mG, a kinetic-to-magnetic energy ratio of $0.5 \pm 0.4$, and our $\mathcal{M}_{\mathrm{A}}$ map, which shows sub-Alfv\'enic flow. The streamer is resolved in both DCN$(J=3\rightarrow2)$ and C$^{18}$O$(J=2\rightarrow1)$ emission, over the velocity range 8.42--9.29~{\kms}, consistent with earlier NOEMA findings \citep{Hsieh2024}.

\item The spiral structure in the dust emission seen earlier by \citet{Tobin2018} and \citet{Diaz-Rodriguez2022} is confirmed here, revealing an asymmetric double spiral.  The dust spiral is spatially well correlated with the streamer seen in molecular line emission and with the magnetic field morphology. 

\item  The polarized emission from dust was interpreted to be the result of magnetically aligned dust grains by comparing the field morphology with the intensity and velocity gradients. The results from the calculation of the AM parameter strongly support this interpretation. To the south of the circumbinary disk, however, we cannot rule out the possibility that we are seeing mechanical alignment from dust grains if the radiation field there is not strong enough to efficiently spin up the grains. 

\end{itemize}

The sub-Alfv\'enic streamer presented here suggests a new role for the magnetic field when gravity dominates, where it acts as a ``guide'' facilitating the infalling of material from the envelope onto the disk.

\appendix
\section{Computing Gradients}
\label{se:appendix_a}

To derive an interpretation for the polarized dust emission seen from the ALMA data, we will compare the orientations of the total intensity continuum emission and the C$^{18}$O velocity field gradients with respect to the magnetic field orientation. The comparison with the polarization position angle is immediate, as they are orthogonal to each other. 

\subsection{Comparing with the intensity gradient}
\label{sse:int_gradient}
\begin{figure*} 
\centering
\includegraphics[width=0.49\hsize]{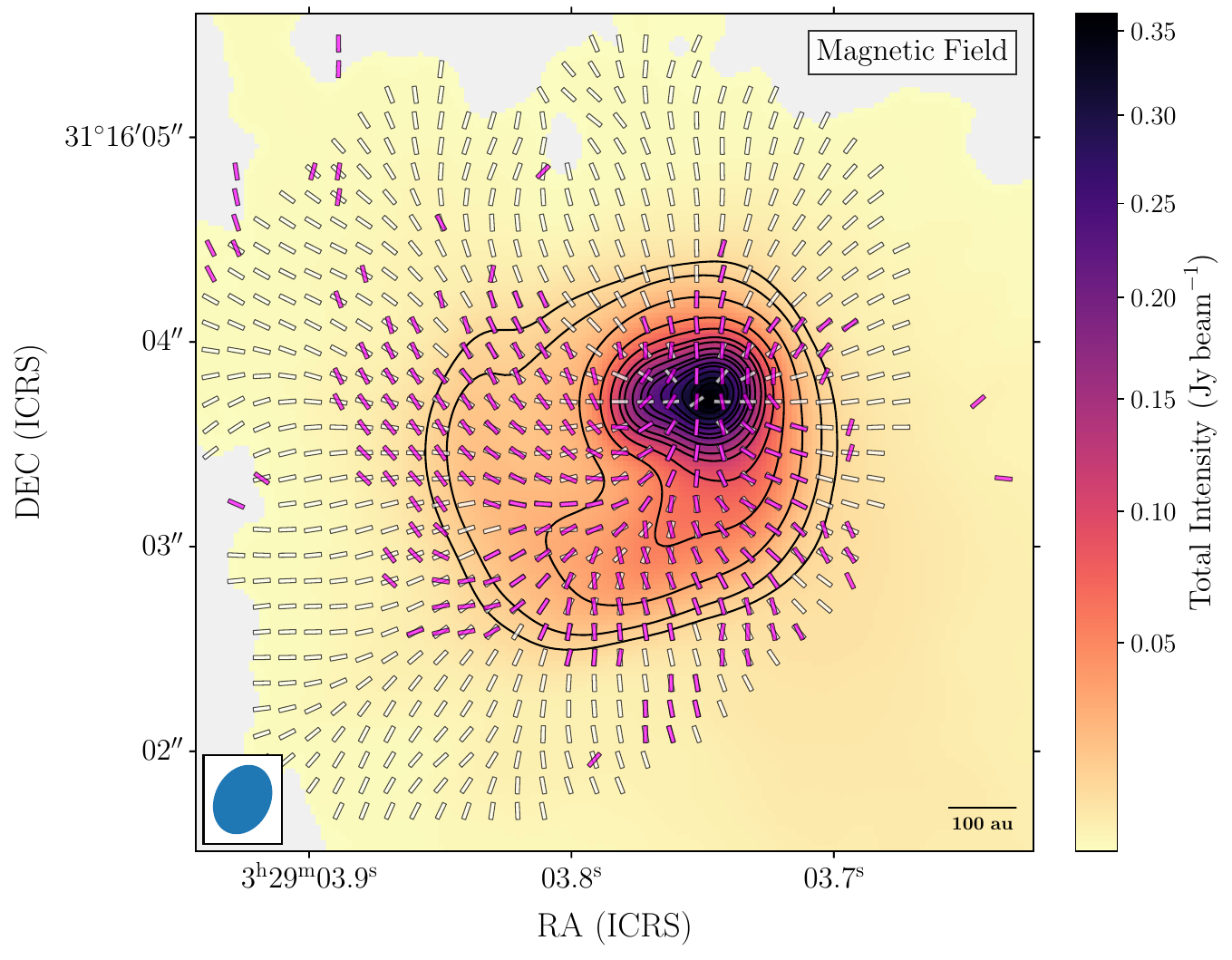}
\includegraphics[width=0.49\hsize]{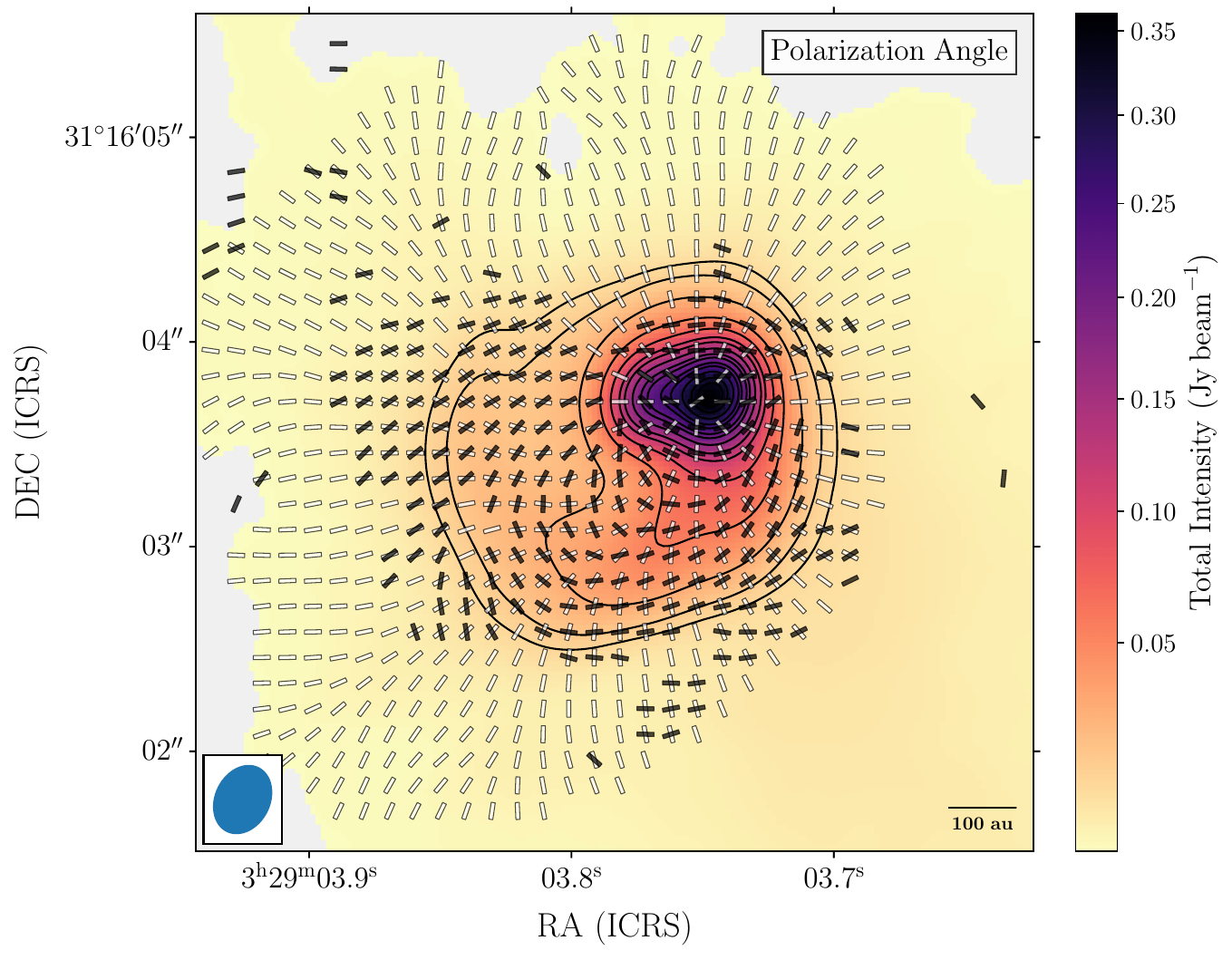}

\caption{{\bf\em Left panel.} The dust emission map shown with robust=0.5 in color scale and in contours using the same levels used in Figure \ref{fig:dust_map}. Superposed in white pseudo-vectors is the intensity gradient. Also, superposed in magenta pseudo-vectors is the inferred magnetic field derived from the polarized emission, also imaged using robust=0.5. {\bf\em Right panel.} Same as the {\em left panel}, but with the polarization position angle as black-pseudo vectors. The beam is depicted as a blue ellipse in the bottom left corner, and the scale is displayed in the bottom right corner of each panel. 
\label{fig:gradient}
}
\bigskip
\end{figure*}

We compute the gradients by following the approach implemented by \citet{Soler2013}, who calculated the column density gradient using Gaussian derivatives. Note that the orientation of the gradient follows the isodensity contours usually plotted in 2-dimensional, or 2D,  maps. The calculation of the discrete derivative becomes equivalent to the convolution with a kernel of a certain size, where one common choice is a $3\times3$ Sobel kernel \citep{Sobel1968}. Using a Gaussian derivative, however, is more convenient because, as differentiation and convolution are commutative, the operations can be interchanged and the computation of the gradient can be performed by applying the Gaussian derivative as a filter\footnote{We implement a version of the Gaussian filter that can handle not-a-number values using the {\em scipy Gaussian2DKernel} object}
\citep[see section 2.1 in ][]{Soler2013}. We define the size of the Gaussian kernel as the number of pixels used to sample the beam when imaging the data, which in our case corresponds to $\sigma_{\mathrm{G}}=8$. Although we are using a Gaussian kernel similar to the beam size, it will still partially smooth the data because the Gaussian kernel is symmetric, and the beam is not. This technique, however, was not applied to the polarization data, and thus any bias introduced by the smoothing will be systematic only to the intensity and the velocity field gradients.
Figure \ref{fig:gradient} shows the comparison between the intensity gradient, the magnetic field, and polarization position angle, where the intensity gradient was calculated over the circumbinary disk up to the 3\,$\sigma$ cutoff in continuum emission only (the bridge to the VLA 3 source was excluded). From visual inspection, the inferred magnetic field appears to agree quite well with the intensity gradient throughout the map, while the polarization does not show the same level of agreement. We will quantitatively compare this later in section \ref{sse:am}.

\subsection{Comparing with the velocity gradient}
\label{sse:v_gradient}

\begin{figure*} 
\centering
\includegraphics[width=0.45\hsize]{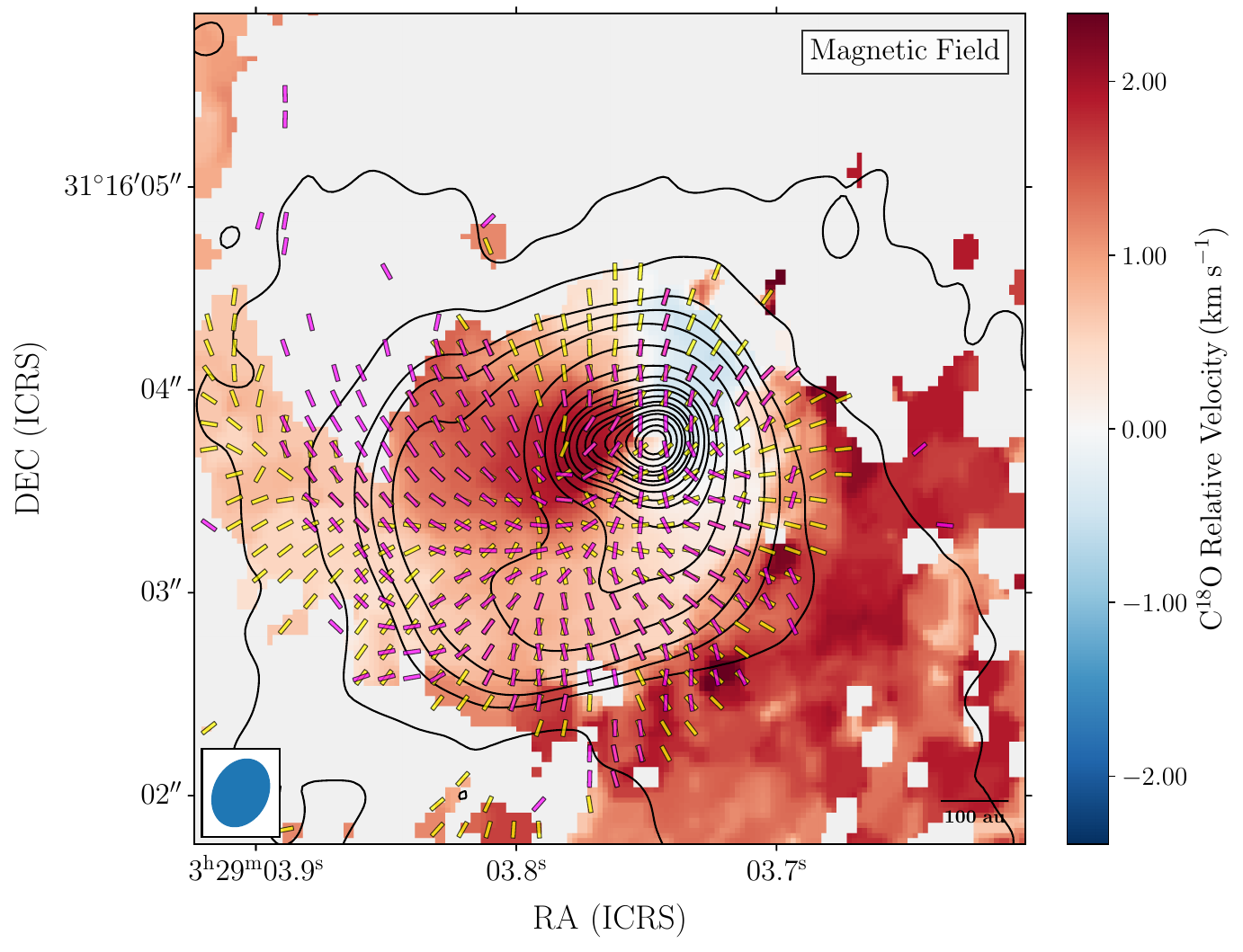}
\includegraphics[width=0.45\hsize]{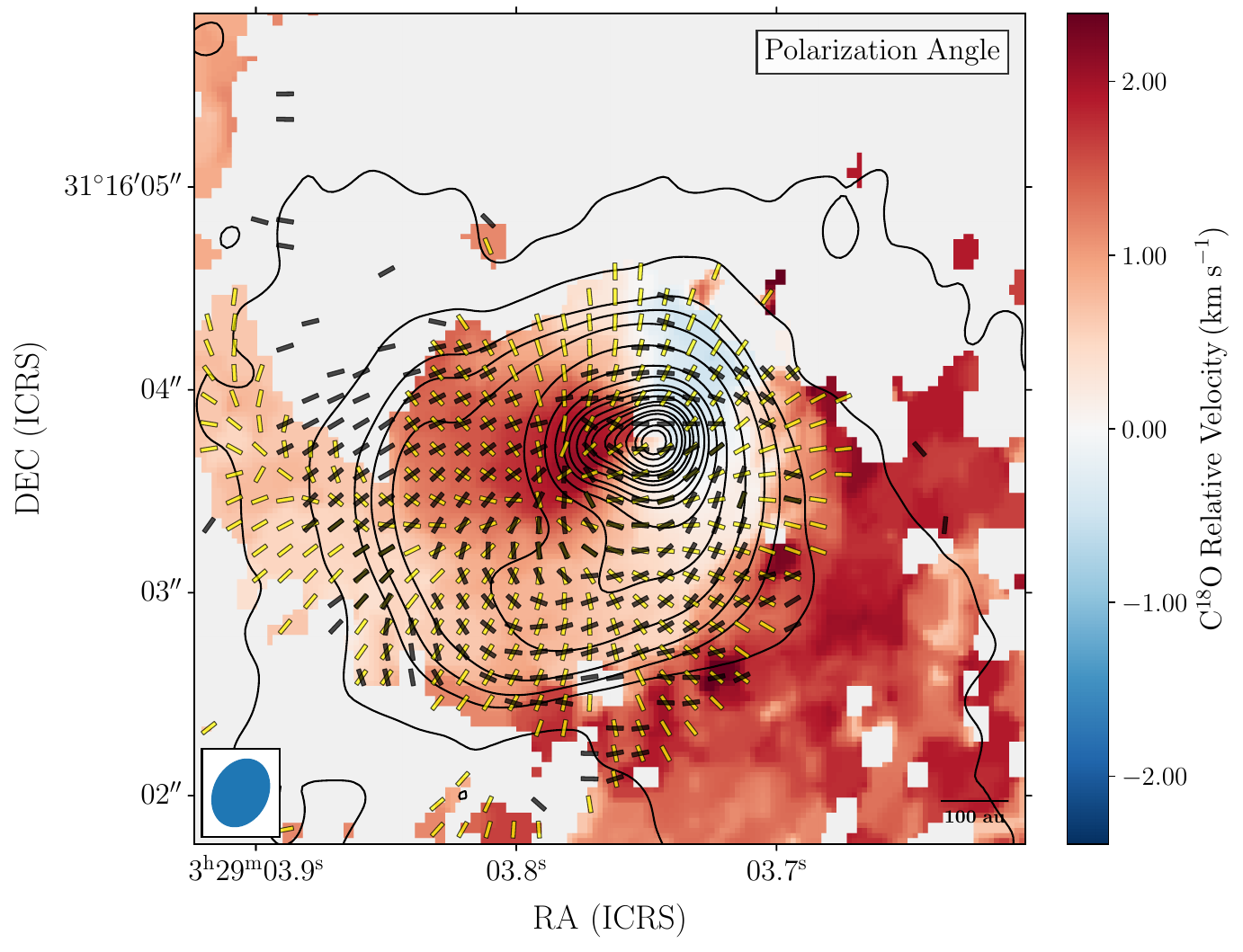}

\caption{{\bf\em Left panel.} The C$^{18}$O moment 1 map shown with robust=0.5 in color scale, where the velocities are relative to the v$_{lsr} = 7.36\, {\kms}$. Superposed in yellow pseudo-vectors is the velocity gradient, which was computed over the moment 1 map. Also, superposed in magenta pseudo-vectors is the inferred magnetic field as derived from the polarized emission, also imaged using robust=0.5. In contours, we show the total intensity of the continuum emission. {\bf\em Right panel.} Same as the {\em left panel}, but with the polarization position angle as black-pseudo vectors.  The beam is depicted as a blue ellipse in the bottom left corner, and the scale is displayed in the bottom right corner of each panel. 
\label{fig:c17o_gradient}
}
\bigskip
\end{figure*}

As previously done in section \ref{sse:int_gradient}, we computed the velocity gradient for the C$^{18}$O velocity field map using the same Gaussian derivative and kernel window (see Figure \ref{fig:c17o_gradient}). This approach is similar to that used by \citet{Liu2023}, but in their work they used a $3\times3$ Sobel kernel and not a Gaussian derivative. From the Figure, the comparison between the velocity gradient and the magnetic field suggests good agreement at red-shifted velocities ($\Delta\ V \gtrsim 0$ \kms\,). The red-shifted positions are consistent with the spiral substructure seen in the dust (see lower panel in Figure \ref{fig:c17o}).
With the blue-shifted velocities, there is also agreement, but in this case, we may be seeing a rotation signature in the gas, and thus, the interpretation is less clear. Nonetheless, the blue-shifted component is rather small in  C$^{18}$O and we ignore it for now.
When inspecting the comparison with the polarization (Figure \ref{fig:c17o_gradient}, lower panel), some pseudo vectors appear to correlate with the velocity gradient to the southern part of the circumbinary disk, which may be a signature of mechanical alignment (see section \ref{sse:mech}).

\section{Alignment Measures}
\label{sse:am}

\begin{figure*} 
\centering
\includegraphics[width=0.45\hsize]{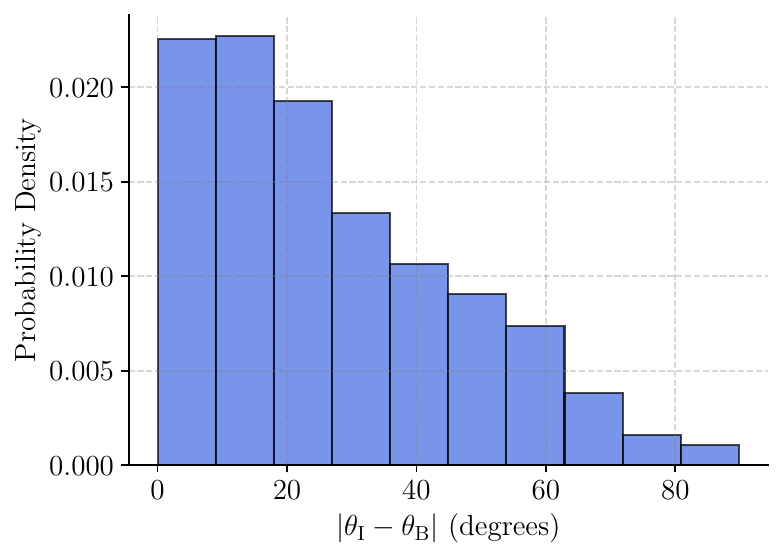}
\includegraphics[width=0.45\hsize]{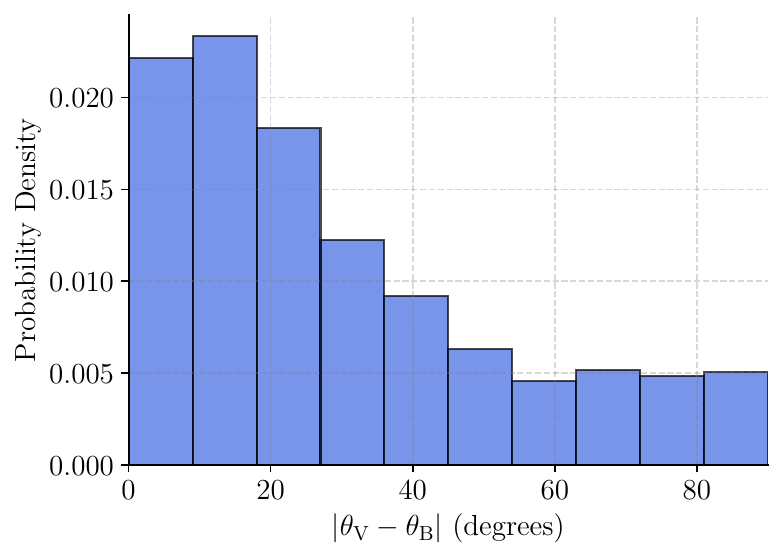}

\caption{{\bf \em Left panel.} The histogram of absolute angle differences between the intensity gradient orientation and the magnetic field is shown here as probability density. {\bf \em Right panel.} Same as the {\em left panel} but for the velocity gradient orientation and the magnetic field. 
\label{fig:angle_diff_hist}
}
\bigskip
\end{figure*}

To quantify the comparison between the inferred magnetic field, the intensity gradient, and the velocity gradient orientations, we computed histograms of the absolute angle differences and of the alignment measure (AM) using 10 equally weighted bins for each distribution.  The AM parameter was introduced by \citet{Gonzales2017} and is calculated as 
\begin{equation}
\mathrm{AM} = \left<\cos{(2\,\theta_{\mathrm{1,2}}}) \right>
\end{equation}
where $\theta_{\mathrm{1,2}} = |\theta_{\mathrm{1}} - \theta_{\mathrm{2}}|$ is the absolute angle difference between either the total intensity gradient and the magnetic field ($\theta_{\mathrm{I,B}}$), and the velocity gradient and the magnetic field ($\theta_{\mathrm{V,B}}$). Values for the alignment measure $\sim 1$  indicate that the average angular difference in the directions of the two vectors across the map is small; if AM $\sim 0$, then there is no alignment between the two vectors. But if AM $\sim -1$,  the two directions tend to be perpendicular to each other \citep{Lazarian2018}. We calculate the AM parameter by directly computing the $\cos{(2\theta_{\mathrm{1,2}}})$ between the two maps and by taking a moving average over the absolute angle difference map following the approach used by \citet{Cortes2024}. In this case, the moving average window is defined by taking the same window used to calculate the gradient using Gaussian derivatives (8 pixels across) to avoid introducing excessive averaging (see Figure \ref{fig:angle_diff_hist}). In this way, we also produce AM maps which we use to aid the interpretation (see section \ref{sse:interpreting}).
 The histograms were computed considering the total extent of the polarized emission and the velocity gradient maps produced using the {\em robust} = 0.5 maps. This robust value appears to provide the best trade-off between the recovery of significant polarized emission and resolution.
The first histogram shows an increasing likelihood of absolute angle difference towards $\theta_{\mathrm{I,B}} = |\theta_{\mathrm{B}} - \theta_{\mathrm{I}}| = 0$, which suggests similar orientations between the intensity gradient and the magnetic field over the region. The comparison with the velocity gradient shows a broader distribution of values, but the highest densities also suggest a similar orientation between the field and the velocity gradient. 
Figure \ref{fig:am_hist}, shows the histogram of the AM values for the same angle differences, which support alignment between the gradients and the inferred magnetic field (for a discussion, see section \ref{sse:interpreting}).

 \subsection{Radiative Alignment}
 \label{sse:radiative_alignment}
Radiative alignment has been proposed as an alternative source of dust polarized emission  \citep{Tazaki2017}. In this scenario, radiative alignment from grains with weak, or no, internal alignment produces a mixed population of grains being aligned with both minor and major axes in the direction of the radiation field \citep{Andersson2022}. The lack of internal alignment comes from
diamagnetic, predominantly carbon grains, which do not develop a magnetic moment when spun up.
In this way, if the radiation field is strong enough and if the grain has good alignment between the angular momentum and the grain axes, the preferred direction might be the radiation field and not the magnetic field.
Two sources where radiative alignment has been proposed to explain the observed polarized emission are $\beta$-Pic \citep{Hull2022} and  IRC+10$^{\circ}$216 \citep{Andersson2022}. The former, a debris disk, shows stacked polarized emission with a polarization fraction profile consistent with radiative alignment, while the latter, a carbon star, shows an almost complete centrosymmetric radial polarization pattern. 
To test radiative alignment in SVS13A, we produced a simple model for the flux emission coming from the binary system, which we assumed to be the source of the radiation flux. The model assumes circular Gaussian profiles for each source at their respective positions, which we used to calculate their gradients. As expected, each gradient is purely radial from each of the VLA 4a and VLA 4b proto-stars.
To compute the combined radiation flux, we did a weighted vector sum of each of the gradients, considering a $1/r^{2}$ pixel dependence and the flux ratio between VLA 4a and VLA 4b, which was done as follows,

\begin{eqnarray}
S = w_{1}\sin{\theta_{1}} + w_{2}w_{f}\sin{\theta_{2}}\\
C = w_{1}\cos{\theta_{1}} + w_{2}w_{f}\cos{\theta_{2}}
\end{eqnarray}

where $w_{1}=1/r_{1}^2$ and $w_{2}=1/r_{2}^2$ are the corresponding radii from VLA 4A and VLA 4B, $w_{f} = 0.08$ is the ratio between the VLA 4a and VLA 4b fluxes \citep{Diaz-Rodriguez2022}, and $\theta_{1}$ and $\theta_{2}$ corresponds to the position angle of the flux model gradient from each source. The combined radiation model angle is calculated as

\begin{equation}
\theta_{c} = \arctan{S/C}
\end{equation}

\begin{figure*} 
\centering
\includegraphics[width=0.99\hsize]{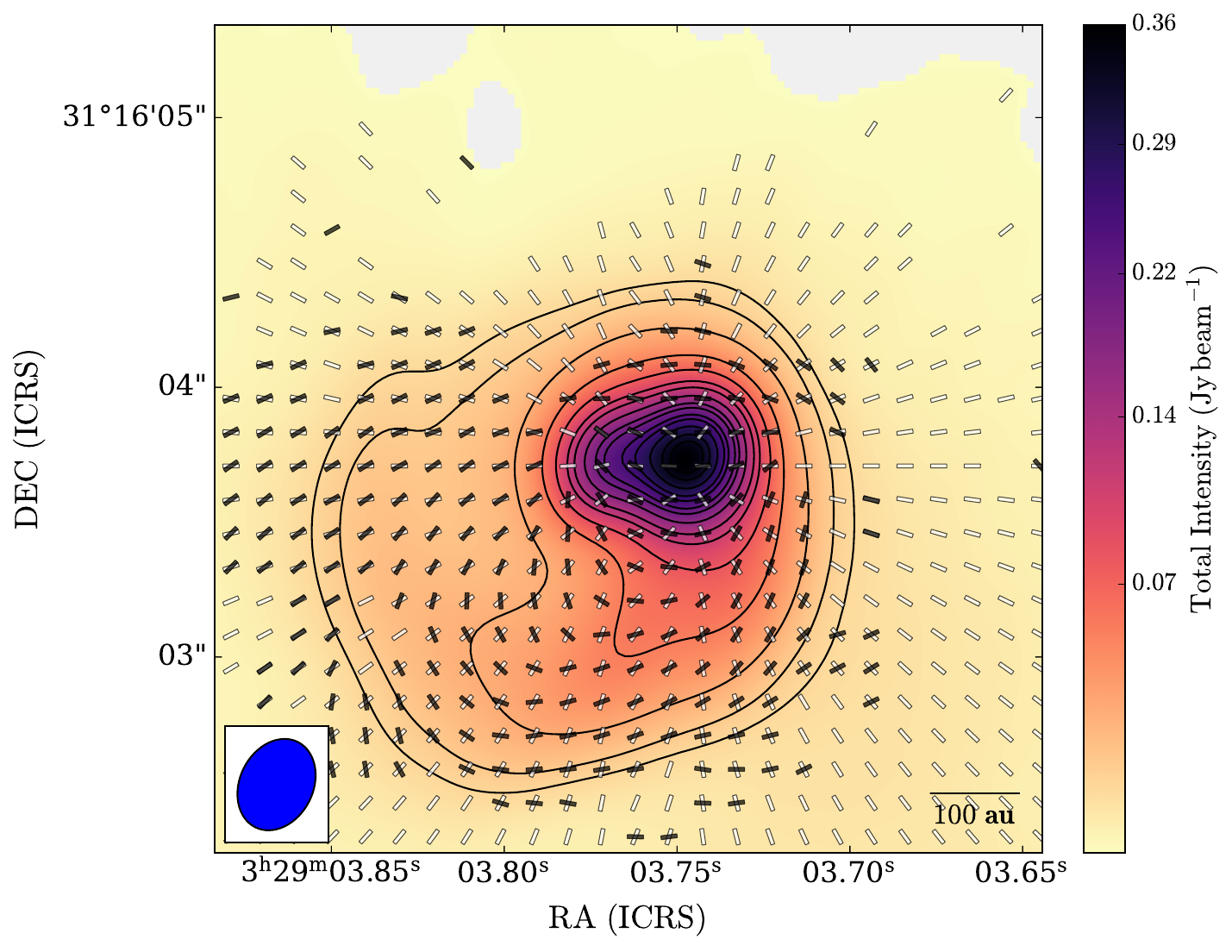}

\caption{A comparison is shown between the polarization position angle map, as black pseudo-vectors, and the radiation field model indicated by white pseudo vectors. In contours and color scale, we show the total intensity of dust emission.
The beam is shown as a blue ellipse in the bottom left corner,  and the scale is shown in the bottom right corner of the map. 
\label{fig:rad}
}
\bigskip
\end{figure*}

Although  VLA 4B  is slightly brighter, with a resolved flux of $110 \pm 11$ mJy with respect to $83 \pm 9$ mJy from VLA 4A at 380 GHz and a projected distance of 90 au between the stars \citep{Diaz-Rodriguez2022},  the combined radiation field is not strictly radial. At longer distances from the binary, the combined radiation field model approaches a radial orientation from the binary's center of mass, modulo optical depth effects. The comparison between the polarization position angle and the radiation field model is shown in Figure \ref{fig:rad}. From visual inspection, the polarization position angle appears not to be aligned with the radiation field throughout most of the map, where a bimodal distribution is not apparent. 
In sources where radiative alignment has been proposed, such as IRC+10$^{\circ}$216, we do not see a bimodal population of polarization angles but rather a completely radial one. This behavior can be explained by adding mechanical alignment by the stellar wind, which orients the grains radially \citep{Andersson2022}. The conditions in SVS13A are completely different from IRC+10$^{\circ}$216, as
the dominant velocity gradient seen at these scales comes from the streamer, where we find that mechanical alignment is not favored. Furthermore, there is no evidence for a single population of just carbon dust grains in SVS13A. It is more likely that the dust grain population around SVS13A is composed of both paramagnetic and diamagnetic dust grains, where the former are more likely to be magnetically aligned if the radiation field is strong \citep{Lazarian2007}. Therefore, we disfavor radiative alignment as the preferred mechanism for the polarized emission that we see in SVS13A.

\section{The effect of Using a Smaller Moving Window}
\label{sse:small_window}

\begin{figure*} 
\centering
\includegraphics[width=0.99\hsize]{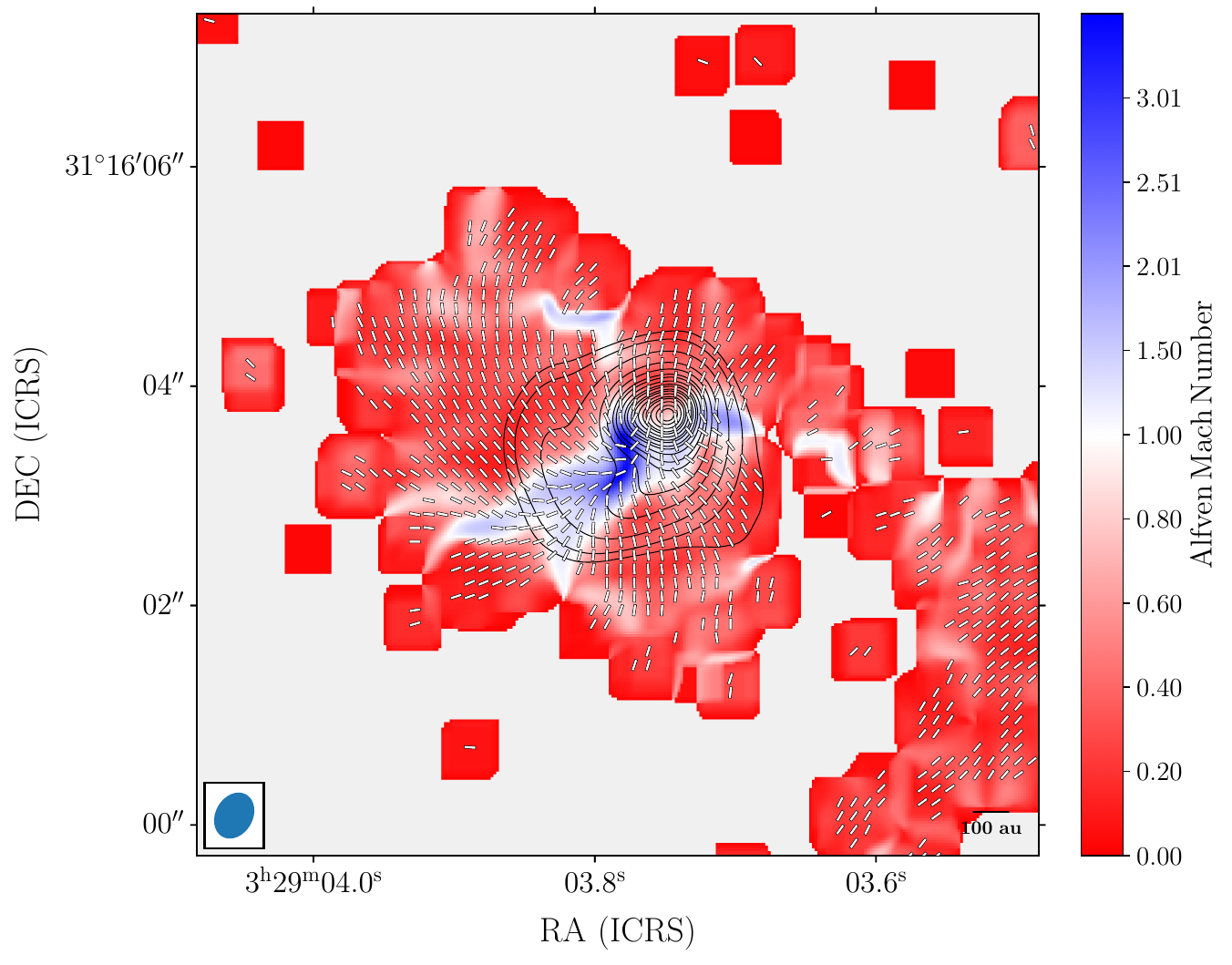}

\caption{The Alfv\'en Mach number from the SVS13A magnetic field map is shown in the color scale. Here, the color white represents the trans-Alfv\'enic value of $\mathcal{M}_{\mathrm{A}} = 1$ while red indicates the sub-Alfv\'enic regime. In this case, the map was produced using a moving window of $0^{\prime\prime}.5$ with an associated error of $10^{\circ}$.
The magnetic field morphology is superposed as white pseudo-vectors plotted every five pixels.
The map was produced from the data imaged using robust=1. The beam is shown as a blue ellipse in the bottom left corner,  and the scale is shown in the bottom right corner of the map. 
\label{fig:am_map_05}
}
\bigskip
\end{figure*}

Using a moving window to estimate the dispersion of the magnetic field lines has the objective of capturing the local perturbations to the field from non-thermal motions. To this effect, it is desirable to use the smallest window possible. A window smaller than the one used in Figure \ref{fig:Alfven_mach_number} introduces a larger error because the number of independent points decreases. 
Figure \ref{fig:am_map_05} shows the $\mathcal{M}_{\mathrm{A}}$ map using a moving window of $0^{\prime\prime}.5$ in size, which corresponds to 1.3 times the beam size for robust = 1, or $0^{\prime\prime}.38$. With such a window size, we would sample the polarization position angle map using $\sim 2\times2$ independent points when Nyquist sampling is considered, yielding an error of about $10^{\circ}$ when we use a population dispersion estimate of  $12^{\circ}$ \citep[see ][ Appendix B]{Cortes2024}. Although using a small window size decreases the spread of larger dispersion values, as seen in the apparent super-Alfv\'enic values shown in Figure \ref{fig:Alfven_mach_number}, the associated error almost doubles from $6^{\circ}$ to $10^{\circ}$, which constitutes an important difference if we consider that the polarization position angle expected error in ALMA's Band 7, within the primary beam area covering the streamer, is approximately $< 2^{\circ}$ \citep[see Figure A.3 in ][]{Hull2020}. 

\begin{acknowledgements}
P.C.C. acknowledges publication and travel support from ALMA, NRAO, and MPE.
L.W.L. acknowledges support from NSF-1910364 and NSF-2307844.
Z.Y.L. is supported in part by NASA 80NSSC20K0533 and NSF AST-2307199.
C.L.H.H. is currently employed by AES US Services.  His contributions were made independently.  When this work was initiated, he was affiliated with the National Astronomical Observatory of Japan, NAOJ Chile, and the ALMA Observatory.
This paper makes use of the following ALMA data: 2017.1.00053.S, 2021.1.00418.S, and 2022.1.00479.S
ALMA is a partnership of ESO (representing its member states), NSF (USA) and NINS (Japan), together with NRC (Canada), MOST and ASIAA (Taiwan), and KASI (Republic of Korea), in cooperation with the Republic of Chile. The Joint ALMA Observatory is operated by ESO, AUI/NRAO and NAOJ.
The National Radio Astronomy Observatory and Green Bank Observatory are facilities of the U.S. National Science Foundation operated under cooperative agreement by Associated Universities, Inc.
\end{acknowledgements}

\bibliography{biblio}
\bibliographystyle{aasjournal}

\end{document}